\documentclass[aps,prd,reprint,nofootinbib,superscriptaddress]{revtex4-2}

\usepackage{amsmath,amssymb,bm}
\usepackage{graphicx}
\usepackage{hyperref}
\usepackage{xcolor}
\usepackage{mathtools}
\hypersetup{
 colorlinks=true,
 linkcolor=blue!45!black,
 citecolor=green!35!black,
 urlcolor=blue!45!black,
 pdftitle={Null-Energy Bounds on Photon-Orbit Instability in Einstein-Gauss-Bonnet Gravity},
 pdfauthor={Sarandas A.; C. Fairoos; C. L. Ahmed Rizwan}
}

\newcommand{\dd}{\mathrm{d}}
\newcommand{\ee}{\mathrm{e}}
\newcommand{\cR}{\mathcal{R}}
\newcommand{\rH}{r_{\mathrm H}}
\newcommand{\rg}{r_\gamma}
\newcommand{\mug}{\mu_\gamma}
\newcommand{\deltag}{\delta_\gamma}
\newcommand{\rhog}{\rho_\gamma}
\newcommand{\prg}{p_{r\gamma}}
\newcommand{\ptg}{p_{t\gamma}}
\newcommand{\kg}{\kappa_\gamma}
\newcommand{\at}{\widetilde{\alpha}}

\begin{document}
\raggedbottom

\title{Null-Energy Bounds on Photon-Orbit Instability in Einstein-Gauss-Bonnet Gravity}

\author{Sarandas A.}
\email{rs\_sarandasa@kannuruniv.ac.in}
\affiliation{Department of Physics, Kannur University, Payyanur, Kerala, India-670327}
\author{C. Fairoos}
\email{fairoos.phy@gmail.com}
\affiliation{Department of Physics, T. K. M. College of Arts and Science, Kollam, Kerala, India-691005}
\author{C. L. Ahmed Rizwan}
\email{ahmedrizwancl@gmail.com}
\affiliation{Department of Physics, Kannur University, Payyanur, Kerala, India-670327}

\begin{abstract}

We derive a local upper bound on the coordinate-time Lyapunov exponent of unstable circular null orbits in static, spherically symmetric, asymptotically flat Einstein--Gauss--Bonnet gravity in $n\geq5$ dimensions with nonnegative Gauss--Bonnet coupling. On the regular Einstein branch, the tangential null energy condition suffices; no additional density-pressure inequality is required for the principal result. The bound retains the local matter-density contribution dictated by the field equations. Using the spherical mass relation, we express the departure of its dimensionless coefficient from the Einstein value $n-3$ in terms of the difference between the local density and the enclosed generalized Misner--Sharp mass per areal volume. This gives a sufficient density-contrast criterion for the Einstein optical bound to remain valid at nonzero coupling. Optical Gaussian curvature and tangential null convergence provide geometrical interpretations that distinguish radial tidal instability from the focusing of a null congruence. Vacuum solutions saturate the bound, while a Maxwell field with nonzero charge produces a strictly positive gap. The smooth Einstein limit and explicit vacuum and charged solutions provide consistency checks. A separate matter condition yields a density-free corollary. We also discuss consequences for strong-deflection lensing and test-field eikonal damping within their respective regimes of validity.
\end{abstract}

\maketitle
\raggedbottom

\section{Introduction}
\label{sec:intro}

Unstable circular null orbits provide a direct link between the local geometry of a black hole and several observable strong-field phenomena.  Photon surfaces admit an invariant geometrical definition \cite{Claudel2001}; in a static, spherically symmetric spacetime, the unstable circular null orbit controlling the capture barrier determines the critical impact parameter separating capture from escape and thereby sets the natural scale of the black-hole shadow \cite{Chandrasekhar1983,PerlickTsupko2022,CunhaHerdeiro2018}.  Light rays with impact parameters close to this critical value can circle the black hole repeatedly before escaping, producing the logarithmic accumulation of images characteristic of strong gravitational lensing \cite{Bozza2002}.  The same orbit also carries a natural dynamical timescale.  A small radial perturbation grows exponentially, and the corresponding Lyapunov exponent $\lambda$ measures the rate at which neighboring null rays separate, and sets the local exponential departure timescale near the photon-sphere \cite{Cardoso2009}.\\

This instability has also become a useful organizing idea in black-hole perturbation theory.  For minimally coupled test fields in the eikonal regime, the real part of a quasinormal frequency is set by the orbital angular frequency, while the damping rate is set by $\lambda$ \cite{Cardoso2009,Decanini2011}; the same pair of scales enters strong-deflection lensing observables \cite{Stefanov2010}.  The relation is compelling because it ties wave propagation to a simple null-geodesic picture.  It is not, however, a license to identify every gravitational quasinormal mode with a null orbit of the background metric.  In Lovelock theories, the different gravitational polarizations may propagate on effective metrics, and the familiar correspondence can fail \cite{Reall2014,KonoplyaStuchlik2017}.  We shall return to this distinction below.\\

Model-independent bounds provide a complementary way to study photon spheres. Rather than specifying a particular metric and evaluating its circular null orbit, one asks what restrictions follow for an entire class of spacetimes from the gravitational field equations and physically motivated conditions on matter. In four-dimensional Einstein gravity, upper bounds on the null-orbit Lyapunov exponent have been obtained in terms of a generalized surface gravity, the critical impact parameter, and local acceleration scales \cite{GalloMadler2025}. Dimension-dependent behavior has also been examined for particular higher-dimensional black holes and fuzzball geometries \cite{Bianchi2020}. For a general static, spherically symmetric, asymptotically flat solution of the $n$-dimensional Einstein equations, the tangential null energy condition implies \cite{Fairoos2026}
\begin{equation}
\lambda\leq\sqrt{n-3}\frac{\kappa_\gamma}{\sqrt{\mu_\gamma}},
\label{eq:Einstein-bound-intro}
\end{equation}
where $\kappa_\gamma$ is the generalized surface gravity and $\mu_\gamma$ is the metric function $\mu(r)$, both evaluated at the circular null orbit.  The same analysis gives the dimensionless optical bound $\lambda R_{\rm sh}\leq\sqrt{n-3}$, where $R_{\rm sh}$ is the shadow radius, equivalently the critical impact parameter measured by an observer at infinity; both inequalities are saturated by the vacuum Schwarzschild-Tangherlini geometry. The inequality for the complete class of Einstein-matter configurations, however, follows from the Einstein equations together with the tangential null energy condition and cannot be inferred from a solution-specific geodesic calculation alone.\\

Einstein-Gauss-Bonnet (EGB) gravity provides a particularly clean setting in which to test whether the Einstein-gravity Lyapunov bound survives higher-curvature corrections. It is the simplest nontrivial Lovelock extension of Einstein gravity and modifies the relation between spacetime curvature and matter while retaining second-order field equations \cite{Lovelock1971}. Moreover, its Einstein branch reduces continuously to general relativity as the Gauss-Bonnet coupling is removed. This continuity allows us to retain the same static, spherically symmetric framework and the same null-geodesic construction while identifying precisely how the field-equation part of the Einstein argument is altered. The Gauss-Bonnet term contributes dynamically for $n\geq5$ and also appears in the low-energy gravitational sector of string-inspired models \cite{Zwiebach1985}; broader aspects of higher-dimensional and Lovelock black holes are reviewed in Refs.~\cite{EmparanReall2008,Charmousis2009}.\\

The static, spherically symmetric sector of EGB gravity is also sufficiently well understood to provide concrete tests of a general bound. Explicit vacuum and electrically charged black-hole solutions are known \cite{BoulwareDeser1985,Wheeler1986,Wiltshire1986}, allowing the general result to be checked in vacuum, in the Einstein limit, and in the presence of a standard matter source. Earlier studies of photon-sphere radius bounds in Einstein and EGB gravity have already shown that higher curvature changes the way in which local matter restrictions are converted into optical statements \cite{Hod2013,GalloVillanueva2015,MaLu2020,Chakraborty2021}. A radius bound, however, does not by itself determine a Lyapunov bound, because the instability also depends on second radial derivatives of the metric \cite{Chakraborty2021}. Recent effective-field-theory analyses likewise demonstrate that higher-curvature corrections can modify photon-surface observables and strong-lensing quantities \cite{Kanai2026}. The remaining task is therefore to determine directly from the EGB field equations how the radial instability of the photon orbit depends on the local geometry and matter distribution.\\

The question addressed here is deliberately local. Given a regular outer photon-sphere in a general static, spherically symmetric EGB solution, which combination of the geometry and the matter distribution determines its instability? We approach this question using the characteristic-function method previously developed in Einstein gravity \cite{GalloMadler2025,Fairoos2026}. The photon-sphere condition and the geometrical relation between the Lyapunov exponent and the radial derivative of the characteristic function depend only on the metric and the null-geodesic equations. The gravitational theory becomes relevant when that derivative is evaluated using the field equations. In EGB gravity, the quadratic curvature term modifies both the purely geometrical contribution and the combination of matter variables that enters the instability. As a result, the tangential null energy condition, which is sufficient for the corresponding matter-independent bound in Einstein gravity, leaves an additional density-dependent contribution. We retain this contribution and make the ordinary-NEC inequality the primary result. Its reformulation through the generalized mass polynomial reveals a density-contrast criterion for retaining the Einstein optical coefficient. Optical curvature and tangential null convergence give complementary interpretations. A tighter inequality that removes the explicit density term is retained as a conditional corollary.

The paper is organized as follows. Section~\ref{sec:setup} introduces the EGB system and specifies the domain of the analysis. The null-geodesic calculation is presented in Sec.~\ref{sec:geodesics}, followed by the derivation of the exact EGB identity in Sec.~\ref{sec:identity}. Section~\ref{sec:bounds} derives the ordinary-NEC bound, its mass-gradient and density-contrast forms, the Einstein limit and a more restrictive corollary. Vacuum solutions and their five-dimensional simplification are discussed in Sec.~\ref{sec:vacuum}, while Sec.~\ref{sec:maxwell} provides a nonvacuum consistency check using a Maxwell field. The physical interpretation and limitations are discussed in Sec.~\ref{sec:discussion}, and the main results are summarized in Sec.~\ref{sec:conclusion}. Supporting derivations and checks are collected in the appendices. We use units in which $G=c=1$ and adopt the metric signature $(-,+,\ldots,+)$.

%%%%%%%%%%%%%%%%%%%%%%%%%%%%%%%%%%%%%%%%%%%%%%%%%%%%%%%%%%%%%%

\section{Einstein-Gauss-Bonnet system}
\label{sec:setup}
Deriving the EGB photon-sphere instability identity requires two distinct ingredients. The null-geodesic equations determine the characteristic function and show that its radial change at the circular orbit governs the coordinate-time Lyapunov exponent. The EGB field equations must then be used to express this radial change in terms of the local geometry and matter distribution. For this reduction, the temporal field equation eliminates the first radial derivative of the metric, the radial field equation relates the characteristic function to the radial pressure, and stress-tensor conservation eliminates the derivative of that pressure. We therefore begin by establishing these equations, defining the combinations in which the Gauss-Bonnet coupling appears, and specifying the regularity conditions required at the circular orbit. Up to boundary terms, which do not affect the bulk field equations used below, the gravity and matter dynamics are described by the EGB action \cite{Lovelock1971,Charmousis2009}
%%%%%%%%%%%%%%%%%%%%%%%%%%%%%%%%%%%%%%%%%%%%%%%%%%%%%%%%%%%%%%
\begin{equation}
 S=\frac{1}{16\pi}\int \dd^n x\sqrt{-g}\,
 \left(R+\alpha\mathcal{L}_{\rm GB}\right)+S_{\rm m},
 \label{eq:action}
\end{equation}
%%%%%%%%%%%%%%%%%%%%%%%%%%%%%%%%%%%%%%%%%%%%%%%%%%%%%%%%%%%%%%
where $n$ is the spacetime dimension, $g$ is the determinant of the metric, $R$ is the Ricci scalar, $\alpha$ is the Gauss-Bonnet coupling, with dimensions of length squared, and $S_{\rm m}$ denotes the action of matter minimally coupled to the metric. The Gauss-Bonnet density is
%%%%%%%%%%%%%%%%%%%%%%%%%%%%%%%%%%%%%%%%%%%%%%%%%%%%%%%%%%%%%%
\begin{equation}
 \mathcal{L}_{\rm GB}
 =R_{\mu\nu\rho\sigma}R^{\mu\nu\rho\sigma}
 -4R_{\mu\nu}R^{\mu\nu}+R^2.
 \label{eq:GB-density}
\end{equation}
%%%%%%%%%%%%%%%%%%%%%%%%%%%%%%%%%%%%%%%%%%%%%%%%%%%%%%%%%%%%%%
Here $R_{\mu\nu\rho\sigma}$ and $R_{\mu\nu}$ are the Riemann and Ricci tensors, respectively. Varying the total action with respect to the metric gives
%%%%%%%%%%%%%%%%%%%%%%%%%%%%%%%%%%%%%%%%%%%%%%%%%%%%%%%%%%%%%%
\begin{equation}
 G_{\mu\nu}+\alpha H_{\mu\nu}=8\pi T_{\mu\nu},
 \label{eq:EGB-field}
\end{equation}
%%%%%%%%%%%%%%%%%%%%%%%%%%%%%%%%%%%%%%%%%%%%%%%%%%%%%%%%%%%%%%
where $G_{\mu\nu}$ is the Einstein tensor, $T_{\mu\nu}$ is the stress tensor obtained from $S_{\rm m}$, and the contribution from the Gauss-Bonnet term is
%%%%%%%%%%%%%%%%%%%%%%%%%%%%%%%%%%%%%%%%%%%%%%%%%%%%%%%%%%%%%%
\begin{align}
 H_{\mu\nu}={}&2RR_{\mu\nu}
 -4R_{\mu\rho}R_\nu{}^\rho
 -4R^{\rho\sigma}R_{\mu\rho\nu\sigma}
 \nonumber\\
 &+2R_\mu{}^{\rho\sigma\eta}R_{\nu\rho\sigma\eta}
 -\frac{1}{2}g_{\mu\nu}\mathcal{L}_{\rm GB}.
 \label{eq:GB-tensor}
\end{align}
%%%%%%%%%%%%%%%%%%%%%%%%%%%%%%%%%%%%%%%%%%%%%%%%%%%%%%%%%%%%%%
For later convenience, we absorb the dimension-dependent factor multiplying the Gauss-Bonnet corrections into
\begin{equation}
%%%%%%%%%%%%%%%%%%%%%%%%%%%%%%%%%%%%%%%%%%%%%%%%%%%%%%%%%%%%%%
 \at=(n-3)(n-4)\alpha.
 \label{eq:alphat}
\end{equation}
%%%%%%%%%%%%%%%%%%%%%%%%%%%%%%%%%%%%%%%%%%%%%%%%%%%%%%%%%%%%%%
We take $n\geq5$, for which the Gauss-Bonnet term contributes dynamically, assume
$\widetilde\alpha\geq0$, and select the Einstein branch, which is continuously
connected to Einstein gravity as $\widetilde\alpha\to0$.

The most general static, spherically symmetric line element in the radial gauge used here is
%%%%%%%%%%%%%%%%%%%%%%%%%%%%%%%%%%%%%%%%%%%%%%%%%%%%%%%%%%%%%%
\begin{equation}
 \dd s^2=-\ee^{-2\delta(r)}\mu(r)\dd t^2
 +\frac{\dd r^2}{\mu(r)}+r^2\dd\Omega_{n-2}^2.
 \label{eq:metric}
\end{equation}
%%%%%%%%%%%%%%%%%%%%%%%%%%%%%%%%%%%%%%%%%%%%%%%%%%%%%%%%%%%%%%
Here $r$ is the areal radius, $\dd\Omega_{n-2}^2$ is the line element on the unit $(n-2)$-sphere, and $\mu(r)$ and $\delta(r)$ are the two independent metric functions. A regular event horizon is located at $r=\rH$, where $\mu(\rH)=0$ and $\delta(\rH)$ remains finite. In the exterior region, $\mu>0$. Asymptotic flatness fixes the normalization of the time coordinate through
%%%%%%%%%%%%%%%%%%%%%%%%%%%%%%%%%%%%%%%%%%%%%%%%%%%%%%%%%%%%%%
\begin{equation}
 \mu(r)\longrightarrow1,
 \qquad
 \delta(r)\longrightarrow0
 \quad (r\longrightarrow\infty).
 \label{eq:asymptotic}
\end{equation}
%%%%%%%%%%%%%%%%%%%%%%%%%%%%%%%%%%%%%%%%%%%%%%%%%%%%%%%%%%%%%%
We allow the matter distribution to be anisotropic and write its mixed components as
%%%%%%%%%%%%%%%%%%%%%%%%%%%%%%%%%%%%%%%%%%%%%%%%%%%%%%%%%%%%%%
\begin{equation}
 T^\mu{}_\nu=\mathrm{diag}(-\rho,p_r,p_t,\ldots,p_t).
 \label{eq:stress}
\end{equation}
%%%%%%%%%%%%%%%%%%%%%%%%%%%%%%%%%%%%%%%%%%%%%%%%%%%%%%%%%%%%%%
The functions $\rho(r)$, $p_r(r)$, and $p_t(r)$ are, respectively, the energy density, radial pressure, and tangential pressure measured by static observers. Minimal coupling of the matter sector implies covariant conservation, $\nabla_\mu T^\mu{}_\nu=0$. Throughout the paper, a prime denotes differentiation with respect to $r$.
The combinations in which the Gauss--Bonnet corrections enter the reduced equations are conveniently expressed as
\begin{equation}
\begin{aligned}
X(r)&=1-\mu(r),
\qquad
z(r)=\frac{\at X(r)}{r^2},\\
\Xi(r)&=1+2z(r)
=1+\frac{2\at X(r)}{r^2}.
\end{aligned}
\label{eq:XD-def}
\end{equation}
The function $X(r)$ measures the departure of $\mu(r)$ from its asymptotic value, while $z(r)$ measures the local strength of the Gauss--Bonnet correction. The dimensionless function $\Xi(r)$ is the normalized derivative of the reduced quadratic Lovelock polynomial:
\begin{equation}
\Xi(r)
=
\frac{1}{r^2}
\left.
\frac{\partial}{\partial X}
\left(r^2X+\at X^2\right)
\right|_r.
\end{equation}
It therefore measures how this polynomial changes with $X$ at fixed areal radius. The function $\Xi$ appears in the denominators when the reduced field equations are solved for radial metric derivatives, so its zeros mark a degeneracy of that reduction. On the regular Einstein branch, $\Xi>0$, and in the Einstein limit $\at\to0$ it approaches unity.

Direct substitution of Eq.~\eqref{eq:metric} into the mixed temporal and radial components of Eq.~\eqref{eq:EGB-field} gives \cite{Wheeler1986,Wiltshire1986,GalloVillanueva2015}
%%%%%%%%%%%%%%%%%%%%%%%%%%%%%%%%%%%%%%%%%%%%%%%%%%%%%%%%%%%%%%
\begin{align}
 8\pi\rho=\frac{n-2}{2r^4}\bigl\{&r^2[rX'+(n-3)X]
 \nonumber\\
 &+\at X[2rX'+(n-5)X]\bigr\}.
 \label{eq:rho-equation}
\end{align}
%%%%%%%%%%%%%%%%%%%%%%%%%%%%%%%%%%%%%%%%%%%%%%%%%%%%%%%%%%%%%%
and
%%%%%%%%%%%%%%%%%%%%%%%%%%%%%%%%%%%%%%%%%%%%%%%%%%%%%%%%%%%%%%
\begin{align}
 8\pi p_r=-\frac{n-2}{2r^4}\bigl\{&
 r^2[rX'+(n-3)X+2r\mu\delta']
 \nonumber\\
 &+\at X[2rX'+(n-5)X+4r\mu\delta']\bigr\}.
 \label{eq:pr-unrearranged}
\end{align}
%%%%%%%%%%%%%%%%%%%%%%%%%%%%%%%%%%%%%%%%%%%%%%%%%%%%%%%%%%%%%%
Adding Eq.~\eqref{eq:rho-equation} and Eq.~\eqref{eq:pr-unrearranged} cancels all terms that do not contain $\delta'$ and gives
%%%%%%%%%%%%%%%%%%%%%%%%%%%%%%%%%%%%%%%%%%%%%%%%%%%%%%%%%%%%%%
\begin{equation}
\delta'=-\frac{8\pi r(\rho+p_r)}
{(n-2)\mu \Xi}.
\label{eq:delta-equation}
\end{equation}
%%%%%%%%%%%%%%%%%%%%%%%%%%%%%%%%%%%%%%%%%%%%%%%%%%%%%%%%%%%%%%
Eq.~\eqref{eq:delta-equation} shows that its radial variation is governed by the matter combination $\rho+p_r$. Stress-tensor conservation supplies a corresponding relation for the radial pressure gradient. Writing the temporal metric coefficient as $A(r)=-g_{tt}$, the radial component of $\nabla_\mu T^\mu{}_\nu=0$ gives
%%%%%%%%%%%%%%%%%%%%%%%%%%%%%%%%%%%%%%%%%%%%%%%%%%%%%%%%%%%%%%
\begin{align}
 p_r'={}&-\frac{A'}{2A}(\rho+p_r)
 +\frac{n-2}{r}(p_t-p_r),
 \nonumber\\
 A(r)={}&\ee^{-2\delta(r)}\mu(r).
 \label{eq:conservation}
\end{align}
%%%%%%%%%%%%%%%%%%%%%%%%%%%%%%%%%%%%%%%%%%%%%%%%%%%%%%%%%%%%%%
At a circular null orbit, the orbital condition fixes $A'/A$, so Eq.~\eqref{eq:conservation} expresses $p_r'$ entirely in terms of the density and the radial and tangential pressures evaluated at the orbit.

The equations above enter the instability calculation in three steps.
First, the temporal field equation determines $\mu'$.
Next, the radial field equation fixes $p_r$ at the circular null orbit.
Finally, stress-tensor conservation expresses $p_r'$ in terms of
$\rho$, $p_r$, and $p_t$, eliminating the pressure derivative
from the calculation. Appendix~\ref{app:reduced-equations} checks the temporal equation using the generalized Lovelock mass polynomial, verifies the expression for $\delta'$ by combining the temporal and radial field equations, and derives the radial conservation law directly from $\nabla_\mu T^\mu{}_r=0$.

We consider an unstable circular null orbit in the regular
static exterior of an Einstein-branch black hole. Its radius
$r_\gamma$ satisfies
\begin{equation}
 r_\gamma>r_H,
 \qquad
 \mu_\gamma>0,
 \qquad
 \Xi_\gamma>0.
 \label{eq:domain}
\end{equation}
The algebraic derivation of the exact instability identity
requires only $\Xi_\gamma\neq0$. Its interpretation as an
upper bound under the tangential null energy condition
requires the positive sign specified above. This sign is
a property of the nondegenerate Einstein branch and does
not require $\mu_\gamma<1$ or nonnegative matter density.

To make the branch restriction explicit, introduce the
generalized EGB mass polynomial \cite{MaedaNozawa2008},
\begin{equation}
 \mathfrak m(r)
 =r^{n-3}X+\at r^{n-5}X^2.
 \label{eq:mass-function}
\end{equation}
Differentiating this expression and using
Eq.~\eqref{eq:rho-equation} gives
\begin{equation}
 \mathfrak m'(r)
 =\frac{16\pi}{n-2}r^{n-2}\rho(r).
 \label{eq:mass-derivative}
\end{equation}
The mass polynomial also implies
\begin{equation}
 \Xi(r)^2
 =1+\frac{4\at\mathfrak m(r)}{r^{n-1}}.
\end{equation}
The Einstein branch selects the positive square root.
We restrict attention to its connected exterior region
where the radicand is strictly positive, excluding points
with $\Xi=0$ at which the two algebraic branches meet.
Thus $\Xi>0$ throughout the stated domain, independently
of the sign of $\rho$. Equivalently, $\Xi(r_H)
=1+2\at/r_H^2>0$, and continuity preserves this sign
throughout an exterior region containing no zero of $\Xi$.

For later comparison, nonnegative density provides a
sufficient condition for the stronger property $\mu<1$.
Indeed,
$\mathfrak m(r_H)=r_H^{n-3}+\at r_H^{n-5}>0$.
If $\rho\geq0$ throughout the exterior, then
Eq.~\eqref{eq:mass-derivative} implies
$\mathfrak m(r)>0$. On the Einstein branch,
\begin{equation}
 X(r)
 =\frac{2\mathfrak m(r)}
 {r^{n-3}[1+\Xi(r)]}>0,
 \qquad \text{if }\mathfrak m(r)>0.
 \label{eq:X-positive}
\end{equation}
This expression also holds in the Einstein limit.
Under this additional density assumption, $\mu<1$,
$z\geq0$, and $\Xi\geq1$. These stronger sign properties
are useful for certain corollaries but are not assumptions
of the principal tangential-NEC bound.

We have now assembled the dynamical information needed to evaluate the orbital instability. The temporal field equation determines the required radial derivative of $\mu$, the radial field equation connects the orbital geometry with the radial pressure, and stress-tensor conservation supplies the derivative of that pressure. With these relations and the relevant regularity conditions in place, we can turn to the null-geodesic problem and identify the geometrical quantity that measures the instability of the circular orbit. The EGB equations established above will then allow that quantity to be written entirely in terms of the local geometry and matter variables at the photon-sphere.

\section{Circular null orbits and their instability}
\label{sec:geodesics}

 The purpose of this section is to identify the condition for a circular null orbit and to determine its radial instability without using the gravitational field equations.  The Lyapunov exponent is obtained from the linearized phase-space dynamics of nearby null geodesics, following the stability-matrix method used in Refs.~\cite{Cardoso2009,GalloMadler2025,Fairoos2026}.  The resulting relation is therefore independent of the EGB equations; those equations will enter only when the geometrical quantity controlling the instability is expressed in terms of the local matter variables.

Spherical symmetry allows every nonradial geodesic to be confined to a fixed orbital plane through the centre. Choose angular coordinates so that its angular trajectory lies on a great circle of the unit $(n-2)$-sphere, parametrized by $\phi$.  Stationarity and rotational symmetry then provide the conserved energy and angular momentum
%%%%%%%%%%%%%%%%%%%%%%%%%%%%%%%%%%%%%%%%%%%%%%%%%%%%%%%%%%%%%%
\begin{equation}
 E=A(r)\dot t,
 \qquad
 L=r^2\dot\phi,
 \label{eq:EL}
\end{equation}
%%%%%%%%%%%%%%%%%%%%%%%%%%%%%%%%%%%%%%%%%%%%%%%%%%%%%%%%%%%%%%
where $A(r)$ is the temporal metric function and the dot denotes differentiation with respect to an affine parameter along the null geodesic.  Substitution of these conserved quantities into the null condition gives the radial first integral
%%%%%%%%%%%%%%%%%%%%%%%%%%%%%%%%%%%%%%%%%%%%%%%%%%%%%%%%%%%%%%
\begin{equation}
 \dot r^2=V(r)
 =\mu(r)\left[\frac{E^2}{A(r)}-\frac{L^2}{r^2}\right].
 \label{eq:radial}
\end{equation}
%%%%%%%%%%%%%%%%%%%%%%%%%%%%%%%%%%%%%%%%%%%%%%%%%%%%%%%%%%%%%%
Here $V(r)$ denotes the radial potential. The conserved quantities $E$ and $L$ are held fixed when its radial derivatives are taken.

A circular null orbit at $r=\rg$ requires both $V(\rg)=0$ and $V'(\rg)=0$.  The first condition gives
%%%%%%%%%%%%%%%%%%%%%%%%%%%%%%%%%%%%%%%%%%%%%%%%%%%%%%%%%%%%%%
\begin{equation*}
 \frac{E^2}{L^2}=\frac{A_\gamma}{\rg^2},
\end{equation*}
%%%%%%%%%%%%%%%%%%%%%%%%%%%%%%%%%%%%%%%%%%%%%%%%%%%%%%%%%%%%%%
 Differentiating the bracket in Eq.~\eqref{eq:radial} and using this ratio in the second circular-orbit condition yields
%%%%%%%%%%%%%%%%%%%%%%%%%%%%%%%%%%%%%%%%%%%%%%%%%%%%%%%%%%%%%%
\begin{equation}
 \left.\frac{A'}{A}\right|_{r=\rg}=\frac{2}{\rg}.
 \label{eq:photon-A}
\end{equation}
%%%%%%%%%%%%%%%%%%%%%%%%%%%%%%%%%%%%%%%%%%%%%%%%%%%%%%%%%%%%%%
Because $A'/A=\mu'/\mu-2\delta'$, this condition can be written conveniently in terms of the photon-sphere characteristic function
%%%%%%%%%%%%%%%%%%%%%%%%%%%%%%%%%%%%%%%%%%%%%%%%%%%%%%%%%%%%%%
\begin{equation}
 \cR(r)=2\mu-r\mu'+2r\mu\delta',
 \qquad
 \cR(\rg)=0.
 \label{eq:R-def}
\end{equation}
%%%%%%%%%%%%%%%%%%%%%%%%%%%%%%%%%%%%%%%%%%%%%%%%%%%%%%%%%%%%%%
Thus the circular null orbit is a zero of $\cR(r)$.  No gravitational field equation has been used in obtaining this result.

The stability of the orbit is determined by linearizing the radial phase-space dynamics with respect to the asymptotically normalized coordinate time $t$.  Let $X^i(t)$ denote a set of radial phase-space variables satisfying \cite{Cardoso2009}
%%%%%%%%%%%%%%%%%%%%%%%%%%%%%%%%%%%%%%%%%%%%%%%%%%%%%%%%%%%%%%
\begin{equation*}
 \frac{\dd X^i}{\dd t}=F^i(X).
\end{equation*}
%%%%%%%%%%%%%%%%%%%%%%%%%%%%%%%%%%%%%%%%%%%%%%%%%%%%%%%%%%%%%%
The circular orbit is a fixed point of this radial system.  If $\delta X^i$ is a small displacement from the fixed point, its linearized evolution is
%%%%%%%%%%%%%%%%%%%%%%%%%%%%%%%%%%%%%%%%%%%%%%%%%%%%%%%%%%%%%%
\begin{equation*}
 \frac{\dd\,\delta X^i}{\dd t}
 =K^i{}_{j}\,\delta X^j,
 \qquad
 K^i{}_{j}
 =\left.\frac{\partial F^i}{\partial X^j}\right|_{\gamma}.
\end{equation*}
%%%%%%%%%%%%%%%%%%%%%%%%%%%%%%%%%%%%%%%%%%%%%%%%%%%%%%%%%%%%%%
The principal Lyapunov exponent measures the exponential growth rate of the unstable perturbation,
%%%%%%%%%%%%%%%%%%%%%%%%%%%%%%%%%%%%%%%%%%%%%%%%%%%%%%%%%%%%%%
\begin{equation*}
 \lambda
 =\lim_{t\rightarrow\infty}
 \frac{1}{t}
 \ln\!\left(
 \frac{\|\delta X(t)\|}{\|\delta X(0)\|}
 \right).
\end{equation*}
%%%%%%%%%%%%%%%%%%%%%%%%%%%%%%%%%%%%%%%%%%%%%%%%%%%%%%%%%%%%%%

For the radial phase space $X^i=(\pi_r,r)$, the canonical radial momentum is
%%%%%%%%%%%%%%%%%%%%%%%%%%%%%%%%%%%%%%%%%%%%%%%%%%%%%%%%%%%%%%
\begin{equation*}
 \pi_r=g_{rr}\dot r=\frac{\dot r}{\mu}.
\end{equation*}
%%%%%%%%%%%%%%%%%%%%%%%%%%%%%%%%%%%%%%%%%%%%%%%%%%%%%%%%%%%%%%
 The null geodesic Hamiltonian at fixed $E$ and $L$ is
%%%%%%%%%%%%%%%%%%%%%%%%%%%%%%%%%%%%%%%%%%%%%%%%%%%%%%%%%%%%%%
\begin{equation*}
 \mathcal H
 =\frac{1}{2}\left[
 -\frac{E^2}{A(r)}
 +\mu(r)\pi_r^2
 +\frac{L^2}{r^2}
 \right].
\end{equation*}
%%%%%%%%%%%%%%%%%%%%%%%%%%%%%%%%%%%%%%%%%%%%%%%%%%%%%%%%%%%%%%
Since $\dot t=E/A$, Hamilton's equations written with respect to the asymptotically normalized coordinate time are
%%%%%%%%%%%%%%%%%%%%%%%%%%%%%%%%%%%%%%%%%%%%%%%%%%%%%%%%%%%%%%
\begin{equation*}
 \frac{\dd r}{\dd t}
 =\frac{A\mu}{E}\pi_r,
 \qquad
 \frac{\dd \pi_r}{\dd t}
 =-\frac{A}{E}\frac{\partial\mathcal H}{\partial r}.
\end{equation*}
%%%%%%%%%%%%%%%%%%%%%%%%%%%%%%%%%%%%%%%%%%%%%%%%%%%%%%%%%%%%%%
At the circular orbit, $\pi_{r\gamma}=0$ and the first radial derivative of the effective potential vanishes. The linearized radial system is therefore off diagonal:
%%%%%%%%%%%%%%%%%%%%%%%%%%%%%%%%%%%%%%%%%%%%%%%%%%%%%%%%%%%%%%
\begin{equation}
 \frac{\dd}{\dd t}
 \begin{pmatrix}
  \delta \pi_r \\[2pt]
  \delta r
 \end{pmatrix}
 =
 \begin{pmatrix}
  0 & K_1 \\
  K_2 & 0
 \end{pmatrix}_{\!\gamma}
 \begin{pmatrix}
  \delta \pi_r \\[2pt]
  \delta r
 \end{pmatrix},
 \qquad
 \lambda_{\pm}=\pm\sqrt{K_1K_2},
 \label{eq:linear-affine}
\end{equation}
%%%%%%%%%%%%%%%%%%%%%%%%%%%%%%%%%%%%%%%%%%%%%%%%%%%%%%%%%%%%%%
where $\lambda_\pm$ are the eigenvalues of the stability matrix. Using the radial first integral $\dot r^2=V(r)$, the two nonzero entries are
%%%%%%%%%%%%%%%%%%%%%%%%%%%%%%%%%%%%%%%%%%%%%%%%%%%%%%%%%%%%%%
\begin{equation*}
 K_1=\frac{A_\gamma}{2E\mu_\gamma}V''(\rg),
 \qquad
 K_2=\frac{A_\gamma\mu_\gamma}{E}.
\end{equation*}
%%%%%%%%%%%%%%%%%%%%%%%%%%%%%%%%%%%%%%%%%%%%%%%%%%%%%%%%%%%%%%
Their product is consequently
%%%%%%%%%%%%%%%%%%%%%%%%%%%%%%%%%%%%%%%%%%%%%%%%%%%%%%%%%%%%%%
\begin{equation*}
 K_1K_2
 =\frac{A_\gamma^2}{2E^2}V''(\rg)
 =\frac{V''(\rg)}{2\dot t_\gamma^2}.
\end{equation*}
%%%%%%%%%%%%%%%%%%%%%%%%%%%%%%%%%%%%%%%%%%%%%%%%%%%%%%%%%%%%%%
For an unstable circular orbit, $K_1K_2>0$, and the coordinate-time Lyapunov exponent is the positive eigenvalue. Thus \cite{Cardoso2009,GalloMadler2025}
%%%%%%%%%%%%%%%%%%%%%%%%%%%%%%%%%%%%%%%%%%%%%%%%%%%%%%%%%%%%%%
\begin{equation}
 \lambda^2=\frac{1}{2\dot t_\gamma^2}
 V''(\rg).
 \label{eq:lambda-def}
\end{equation}
%%%%%%%%%%%%%%%%%%%%%%%%%%%%%%%%%%%%%%%%%%%%%%%%%%%%%%%%%%%%%%
The factor $\dot t_\gamma^{-2}$ converts the affine-parameter instability rate into the rate measured with respect to the asymptotically normalized coordinate time. In the convention $\dot r^2=V(r)$ used here, radial instability corresponds to $V''(\rg)>0$.

It remains to express the second derivative in Eq.~\eqref{eq:lambda-def} in terms of the characteristic function.  Introduce,
%%%%%%%%%%%%%%%%%%%%%%%%%%%%%%%%%%%%%%%%%%%%%%%%%%%%%%%%%%%%%%
\begin{equation*}
 f(r)=\frac{A(r)}{r^2}.
\end{equation*}
%%%%%%%%%%%%%%%%%%%%%%%%%%%%%%%%%%%%%%%%%%%%%%%%%%%%%%%%%%%%%%
Eq.~\eqref{eq:radial} can then be written as
%%%%%%%%%%%%%%%%%%%%%%%%%%%%%%%%%%%%%%%%%%%%%%%%%%%%%%%%%%%%%%
\begin{equation}
 V(r)=\ee^{2\delta(r)}
 \left[E^2-L^2f(r)\right].
 \label{eq:Vr-f}
\end{equation}
%%%%%%%%%%%%%%%%%%%%%%%%%%%%%%%%%%%%%%%%%%%%%%%%%%%%%%%%%%%%%%
The two circular-orbit conditions are equivalent to
%%%%%%%%%%%%%%%%%%%%%%%%%%%%%%%%%%%%%%%%%%%%%%%%%%%%%%%%%%%%%%
\begin{equation*}
 E^2=L^2f_\gamma,
 \qquad
 f'_\gamma=0.
\end{equation*}
%%%%%%%%%%%%%%%%%%%%%%%%%%%%%%%%%%%%%%%%%%%%%%%%%%%%%%%%%%%%%%
Differentiating Eq.~\eqref{eq:Vr-f}, one obtains
%%%%%%%%%%%%%%%%%%%%%%%%%%%%%%%%%%%%%%%%%%%%%%%%%%%%%%%%%%%%%%
\begin{equation}
 V''(\rg)=-L^2\ee^{2\deltag}f''_\gamma.
 \label{eq:Vr-second}
\end{equation}
%%%%%%%%%%%%%%%%%%%%%%%%%%%%%%%%%%%%%%%%%%%%%%%%%%%%%%%%%%%%%%
The relation between $f(r)$ and the characteristic function follows directly from their definitions.  Differentiating $f$, using $A=\ee^{-2\delta}\mu$, and collecting the resulting terms gives
%%%%%%%%%%%%%%%%%%%%%%%%%%%%%%%%%%%%%%%%%%%%%%%%%%%%%%%%%%%%%%
\begin{equation}
 f'(r)=-\frac{\ee^{-2\delta(r)}}{r^3}\cR(r),
 \qquad
 f''_\gamma
 =-\frac{\ee^{-2\deltag}}{\rg^3}\cR'_\gamma.
 \label{eq:fprime-R}
\end{equation}
%%%%%%%%%%%%%%%%%%%%%%%%%%%%%%%%%%%%%%%%%%%%%%%%%%%%%%%%%%%%%%
In the second relation, terms proportional to $\cR(\rg)$ have vanished because the circular-orbit condition requires $\cR(\rg)=0$.  Combining Eq.~\eqref{eq:Vr-second} and Eq.~\eqref{eq:fprime-R} give
%%%%%%%%%%%%%%%%%%%%%%%%%%%%%%%%%%%%%%%%%%%%%%%%%%%%%%%%%%%%%%
\begin{equation*}
 V''(\rg)=\frac{L^2}{\rg^3}\cR'_\gamma.
\end{equation*}
%%%%%%%%%%%%%%%%%%%%%%%%%%%%%%%%%%%%%%%%%%%%%%%%%%%%%%%%%%%%%%
Moreover, Eq.~\eqref{eq:EL} and Eq.~\eqref{eq:photon-A}, together with $\frac{E^2}{L^2}=\frac{A_\gamma}{\rg^2}$, imply
%%%%%%%%%%%%%%%%%%%%%%%%%%%%%%%%%%%%%%%%%%%%%%%%%%%%%%%%%%%%%%
\begin{equation*}
 \dot t_\gamma^2
 =\frac{E^2}{A_\gamma^2}
 =\frac{L^2}{\rg^2A_\gamma}.
\end{equation*}
%%%%%%%%%%%%%%%%%%%%%%%%%%%%%%%%%%%%%%%%%%%%%%%%%%%%%%%%%%%%%%
Substitution of these two relations into Eq.~\eqref{eq:lambda-def} yields
%%%%%%%%%%%%%%%%%%%%%%%%%%%%%%%%%%%%%%%%%%%%%%%%%%%%%%%%%%%%%%
\begin{equation}
 \lambda^2
 =\frac{\ee^{-2\deltag}\mug}{2\rg}\,
 \cR'_\gamma.
 \label{eq:lambda-R}
\end{equation}
%%%%%%%%%%%%%%%%%%%%%%%%%%%%%%%%%%%%%%%%%%%%%%%%%%%%%%%%%%%%%%
Eq.~\eqref{eq:lambda-R} shows that the radial instability of the circular null orbit is controlled by the slope of the characteristic function at its zero.  Since the prefactor is positive in the static exterior, an unstable orbit has $\cR'_\gamma>0$.  Up to this point, we have used only the metric and the null-geodesic equations to obtain an expression for a stability-measure . 
%%%%%%%%%%%%%%%%%%%%%%%%%%%%%%%%%%%%%%%%%%%%%%%%%%%%%%%%%%%%%%
\section{Exact EGB identity at the photon-sphere}
\label{sec:identity}

We now use the EGB field equations to replace the radial derivative of the characteristic function in the expression of the Lyapunov exponent Eq. (\ref{eq:lambda-R}) with quantities describing the local geometry and matter distribution at the photon-sphere.  Substituting Eq.~\eqref{eq:R-def} into Eq.~\eqref{eq:pr-unrearranged}, and collecting the terms proportional to $\cR$, follows
%%%%%%%%%%%%%%%%%%%%%%%%%%%%%%%%%%%%%%%%%%%%%%%%%%%%%%%%%%%%%%
\begin{align}
 \frac{8\pi p_r}{n-2}={}&
 -\frac{\cR-(n-1)\mu+(n-3)}{2r^2}
 \nonumber\\
 &+\frac{\at(\mu-1)}{2r^4}
 [2\cR-(n-1)\mu+(n-5)].
 \label{eq:pr-R}
\end{align}
%%%%%%%%%%%%%%%%%%%%%%%%%%%%%%%%%%%%%%%%%%%%%%%%%%%%%%%%%%%%%%
Thus Eq.~\eqref{eq:pr-R} is not an additional field equation; it is simply the radial component written in variables adapted to the orbit. Solving Eq.~\eqref{eq:pr-R} for $\cR$, differentiating the resulting expression, and then imposing $\cR_\gamma=0$ introduces $\mu'_\gamma$, $p_{r\gamma}$, and $p'_{r\gamma}$. These quantities are eliminated in a definite order. The temporal field in Eq.~\eqref{eq:rho-equation} determines $\mu'_\gamma$, while the radial equation in Eq.~\eqref{eq:pr-R}, evaluated at $\cR_\gamma=0$, determines $p_{r\gamma}$. Finally, the conservation law, together with the circular-orbit condition in Eq.~\eqref{eq:photon-A}, gives
%%%%%%%%%%%%%%%%%%%%%%%%%%%%%%%%%%%%%%%%%%%%%%%%%%%%%%%%%%%%%%
\begin{equation}
 p'_{r\gamma}
 =\frac{1}{\rg}
 \left[-\rhog+(n-2)\ptg-(n-1)\prg\right].
 \label{eq:conservation-gamma}
\end{equation}
%%%%%%%%%%%%%%%%%%%%%%%%%%%%%%%%%%%%%%%%%%%%%%%%%%%%%%%%%%%%%%
Carrying out these substitutions and collecting the purely geometrical and matter contributions gives the exact result (see Appendix~\ref{app:algebra}),
%%%%%%%%%%%%%%%%%%%%%%%%%%%%%%%%%%%%%%%%%%%%%%%%%%%%%%%%%%%%%%
\begin{align}
 \cR'_\gamma={}&\frac{2}{\rg}F_n(z_\gamma)
 \nonumber\\
 &-\frac{16\pi\rg}{\Xi_\gamma^2}
 \left[\Xi_\gamma(\rhog+\ptg)-\frac{4\at}{(n-2) \rg^2}\rhog\right],
 \label{eq:Rprime-master}
\end{align}
%%%%%%%%%%%%%%%%%%%%%%%%%%%%%%%%%%%%%%%%%%%%%%%%%%%%%%%%%%%%%%
where, 
%%%%%%%%%%%%%%%%%%%%%%%%%%%%%%%%%%%%%%%%%%%%%%%%%%%%%%%%%%%%%%
\begin{equation}
 F_n(z)=(n-3)-\frac{2(n-1)z(1+z)}{(1+2z)^2}.
 \label{eq:Fn}
\end{equation}
%%%%%%%%%%%%%%%%%%%%%%%%%%%%%%%%%%%%%%%%%%%%%%%%%%%%%%%%%%%%%%
Combining Eq.~\eqref{eq:lambda-R} and Eq.~\eqref{eq:Rprime-master} yields the central identity
%%%%%%%%%%%%%%%%%%%%%%%%%%%%%%%%%%%%%%%%%%%%%%%%%%%%%%%%%%%%%%
\begin{equation}
 \begin{aligned}
 \lambda^2={}&\frac{\ee^{-2\deltag}\mug}{\rg^2}F_n(z_\gamma)
 \\
 &-\frac{8\pi\ee^{-2\deltag}\mug}{\Xi_\gamma^2}
 \left[\Xi_\gamma(\rhog+\ptg)-\frac{4\at}{(n-2)\rg^2}\rhog\right].
 \end{aligned}
 \label{eq:lambda-exact}
\end{equation}
%%%%%%%%%%%%%%%%%%%%%%%%%%%%%%%%%%%%%%%%%%%%%%%%%%%%%%%%%%%%%%
This is an equality, not a bound.  It also makes the physics of the higher-curvature correction transparent.  The first line changes the geometrical response of the orbit, while the second line changes how local matter focuses or defocuses neighboring null trajectories.  Radial pressure has disappeared because the photon-sphere equation and stress-tensor conservation tie it to the other local data.  No energy condition has been used.

The Einstein limit is immediate.  As $\at\to0$, $z_\gamma\to0$, $\Xi_\gamma\to 1$, and $F_n(0)=n-3$.  Eq.~\eqref{eq:Rprime-master} becomes
%%%%%%%%%%%%%%%%%%%%%%%%%%%%%%%%%%%%%%%%%%%%%%%%%%%%%%%%%%%%%%
\begin{equation}
 \cR'_\gamma\longrightarrow
 \frac{2(n-3)}{\rg}-16\pi\rg(\rhog+\ptg),
 \label{eq:Einstein-Rprime}
\end{equation}
%%%%%%%%%%%%%%%%%%%%%%%%%%%%%%%%%%%%%%%%%%%%%%%%%%%%%%%%%%%%%%
which is precisely the higher-dimensional Einstein identity \cite{Fairoos2026}.
%%%%%%%%%%%%%%%%%%%%%%%%%%%%%%%%%%%%%%%%%%%%%%%%%%%%%%%%%%%%%%
\section{The tangential-NEC bound and its Einstein limit}
\label{sec:bounds}

The exact identity Eq.~\eqref{eq:lambda-exact} yields a useful upper bound without requiring the full matter bracket to be nonnegative. The relevant assumption is the tangential null energy condition (NEC),
\begin{equation}
 \rhog+\ptg\geq0.
 \label{eq:tNEC}
\end{equation}
 It does not require $\rhog\geq0$; nonnegative density is a separate part of the weak energy condition \cite{MaedaMartinez2020}. We retain $\Xi_\gamma>0$ and the regular static-orbit assumptions of Sec.~\ref{sec:setup}.

\subsection{A sharp local inequality under the ordinary NEC}

Since $\Xi_\gamma(\rhog+\ptg)\geq0$, discarding its nonpositive contribution in Eq.~\eqref{eq:lambda-exact} gives
\begin{equation}
 \lambda^2\leq\frac{\ee^{-2\deltag}\mug}{\rg^2}
 \left[F_n(z_\gamma)+
 \frac{32\pi\at\rhog}{(n-2)(1+2z_\gamma)^2}\right].
 \label{eq:NEC-bound}
\end{equation}
This is our principal bound. It uses the ordinary tangential NEC and retains the density dependence dictated by the EGB equations. Note that, for positive Gauss–Bonnet coupling, positive density increases this bound at fixed local geometric factors. This does not imply that adding matter increases the actual instability: the exact result also depends on the tangential pressure, and changing the matter distribution generally changes the geometry and the photon-orbit radius.

To display the content of the bound, introduce
\begin{align}
 B(r)&=F_n(z(r))+\frac{32\pi\at\rho(r)}{(n-2)\Xi(r)^2}.
 \label{eq:B-def}
\end{align}
Only $B_\gamma$ will be used as an orbital bound. The null condition and the conserved energy and angular momentum give
\begin{equation}
 \Omega_\gamma=\frac{\dd\phi}{\dd t}
 =\frac{\sqrt{A_\gamma}}{\rg},\qquad
 b_\gamma=\frac{L}{E}=\frac{\rg}{\sqrt{A_\gamma}}
 =\Omega_\gamma^{-1}.
 \label{eq:shadow-def}
\end{equation}

Here $\Omega_\gamma$ is the orbital angular frequency measured with respect to the asymptotically normalized time, and $b_\gamma$ is the corresponding impact parameter. If this orbit sets the critical impact parameter separating light
rays arriving from infinity that are captured from those that
escape, then $b_\gamma$ equals the shadow radius $R_{\rm sh}$
measured in impact-parameter units by an observer at infinity
\cite{PerlickTsupko2022}. If several unstable circular null orbits exist, the orbit determining this threshold must be identified from the radial potential. The inequalities derived below apply to the selected unstable orbit, and yield shadow bounds whenever $b_\gamma=R_{\rm sh}$.

Define a dimensionless instability parameter as,
\begin{equation}
 \eta_\gamma=\lambda/\Omega_\gamma=\lambda b_\gamma.
 \label{eq:eta-def}
\end{equation}
Dividing Eq.~\eqref{eq:lambda-exact} by $\Omega_\gamma^2=A_\gamma/\rg^2$ gives the particularly simple exact identity
\begin{equation}
 \eta_\gamma^2=B_\gamma
 -\frac{8\pi\rg^2}{\Xi_\gamma}(\rhog+\ptg).
 \label{eq:shadow-exact}
\end{equation}
The ratio $\eta_\gamma$ measures radial instability per unit orbital angle: the growing linear mode satisfies $\Delta r\propto e^{\eta_\gamma\Delta\phi}$. Thus $2\pi\eta_\gamma$ is the exponential growth exponent per revolution. Since the coefficient of the NEC combination is strictly positive,
\begin{equation}
 0<\eta_\gamma^2\leq B_\gamma,\qquad
 \lambda\leq\Omega_\gamma\sqrt{B_\gamma}.
 \label{eq:NEC-optical}
\end{equation}
Strict instability ensures $B_\gamma>0$, so the square root requires no further assumption. Equality holds if and only if $\rhog+\ptg=0$. In particular, vacuum Einstein-branch black holes saturate the bound. This establishes sharpness in the sense of actual saturating solutions in the stated class, rather than claiming that every prescribed nonzero density is realizable by a global saturating black hole.

The bound also admits an interpretation in terms of the acceleration required to remain static outside the black hole. A static observer follows an integral curve of the timelike Killing vector $\xi^\mu=(\partial_t)^\mu$, with four-velocity $u^\mu=\xi^\mu/\sqrt{A}$. Such an observer must accelerate to maintain a fixed radial position. The acceleration $a^\mu=u^\nu\nabla_\nu u^\mu$ has radial component and magnitude
\begin{equation}
a^r=\frac{\mu A'}{2A},
\qquad
a=\sqrt{a_\mu a^\mu}
=\frac{\sqrt{\mu}\,|A'|}{2A}.
\tag{44}
\end{equation}
Here $a$ is the proper acceleration measured by an accelerometer carried by the observer.

To associate this local acceleration with the time normalization at infinity, multiply its outward orthonormal component by the redshift factor $\sqrt A$. This defines the generalized surface gravity \cite{AbreuVisser2010,GalloMadler2025},
\begin{equation}
\begin{aligned}
\kappa(r)
&\equiv \sqrt A\,a_\mu e_{\hat r}^{\mu}
=\sqrt{\mu}\,\frac{\mathrm d\sqrt A}{\mathrm dr}\\
&=\frac12e^\delta A'
=e^{-\delta}\left(\frac{\mu'}2-\mu\delta'\right),
\end{aligned}
\tag{45}
\end{equation}
where $e_{\hat r}^{\mu}=\sqrt{\mu}\,(\partial_r)^\mu$ is the outward unit radial vector. Thus $\kappa(r)$ represents the redshifted acceleration needed to support a static observer at radius $r$. We retain its sign to distinguish outward from inward acceleration; wherever $A'>0$, it satisfies $\kappa=\sqrt A\,a>0$.

The designation ``generalized surface gravity'' reflects that this quantity is defined throughout the static exterior. At a regular Killing horizon, its limiting value reproduces the usual horizon surface gravity,
\begin{equation}
\kappa_H=\lim_{r\to r_H^+}\kappa(r)
=\frac12e^{-\delta_H}\mu'_H.
\notag
\end{equation}
For a nonextremal horizon, the locally measured acceleration diverges as the horizon is approached, while the redshift factor vanishes, leaving this finite limit.

We now evaluate the same quantity at the photon sphere, defining $\kappa_\gamma\equiv\kappa(r_\gamma)$. The circular-null-orbit condition $A'_\gamma/A_\gamma=2/r_\gamma$ gives
\begin{equation}
\kappa_\gamma
=\frac{e^{-\delta_\gamma}\mu_\gamma}{r_\gamma},
\qquad
\Omega_\gamma=\frac{\kappa_\gamma}{\sqrt{\mu_\gamma}}.
\label{kappa_gamma}
\end{equation}
In particular, the supporting acceleration at the orbit is outward, with magnitude $a_\gamma=\sqrt{\mu_\gamma}/r_\gamma$, and $\kappa_\gamma=\sqrt{A_\gamma}\,a_\gamma$. Eq. (\ref{kappa_gamma}) therefore relates the orbital angular frequency to the redshifted acceleration of a static observer at the same radius.

Substituting this relation into Eq. (\ref{eq:NEC-optical}) yields
\begin{equation}
\lambda\leq
\frac{\kappa_\gamma}{\sqrt{\mu_\gamma}}\sqrt{B_\gamma}.
\label{47}
\end{equation}
This is the acceleration form of the principal NEC bound. The acceleration scale entering it is evaluated at the photon sphere, whereas the horizon surface gravity is evaluated at $r_H$. Consequently, Eq. (\ref{47}) does not assert a bound on $\lambda$ in terms of $\kappa_H$.

\subsection{Mass variation and density contrast}

The bound coefficient $B(r)$ contains both the geometrical term $F_n(z(r))$ and an explicit contribution from the local matter density, as shown in Eq.~\eqref{eq:B-def}. Using the mass relation, we can combine these contributions to express $B(r)-(n-3)$ in terms of the radial derivative of $\Xi(r)$. The mass polynomial Eq.~\eqref{eq:mass-function} implies
\begin{equation}
 \Xi^2=1+\frac{4\at\mathfrak m}{r^{n-1}},\qquad
 z(1+z)=\frac{\at\mathfrak m}{r^{n-1}}.
 \label{eq:Xi-mass}
\end{equation}
Taking its radial derivative on the branch with $\Xi>0$ yields
\begin{equation}
 \frac{r\Xi'}{\Xi}
 =\frac{2\at\,[r\mathfrak m'-(n-1)\mathfrak m]}
 {r^{n-1}\Xi^2}.
 \label{eq:Xi-gradient}
\end{equation}
Using the expression for $F_n$ in Eq.~\eqref{eq:Fn} ,Eqs.~\eqref{eq:mass-derivative}, and \eqref{eq:Xi-mass} in Eq.~\eqref{eq:B-def} therefore gives
\begin{equation}
 B(r)=n-3+\frac{r\Xi'}{\Xi}.
 \label{eq:B-gradient}
\end{equation}
No weak-coupling approximation or equation of state has been used.

The normalized generalized Misner-Sharp mass and the associated areal-volume density scale are \cite{MaedaNozawa2008}
\begin{align}
 M_{\rm MS}(r)&=\frac{(n-2)\Omega_{n-2}}{16\pi}\mathfrak m(r),\\
 V_{\rm ar}(r)&=\frac{\Omega_{n-2}}{n-1}r^{n-1},\\
 \bar\rho(r)&=\frac{M_{\rm MS}(r)}{V_{\rm ar}(r)}
 =\frac{(n-1)(n-2)\mathfrak m(r)}{16\pi r^{n-1}}.
 \label{eq:mean-density}
\end{align}
Here $\Omega_{n-2}$ denotes the area of the unit $(n-2)$-sphere. The quantity $\bar\rho(r)$ provides a density scale associated with the total mass enclosed within the sphere of areal radius $r$. For a black hole, this mass contains both a horizon contribution and the contribution from matter between the horizon and that sphere. Indeed, integrating Eq. (\ref{eq:mass-derivative}) gives
\begin{equation}
\mathfrak m(r)
=
\mathfrak m(r_H)
+\frac{16\pi}{n-2}
\int_{r_H}^{r}\rho(s)s^{n-2}\,\mathrm ds.
\tag{55}
\end{equation}
Thus $\bar\rho$ remains nonzero even in a vacuum exterior. It represents enclosed mass per areal volume, rather than a proper-volume average of the exterior matter density.

Using the definition of $\bar\rho$ in Eq. (\ref{eq:mean-density}), the mass-gradient expression in Eq. (\ref{eq:Xi-gradient}), and Eq. (\ref{eq:B-gradient}), the coefficient entering the NEC bound becomes
\begin{equation}
B(r)
=
n-3+
\frac{32\pi\widetilde\alpha}
{(n-2)\Xi(r)^2}
\left[\rho(r)-\bar\rho(r)\right].
\label{eq:B-contrast}
\end{equation}
This form shows how the local matter density and the enclosed mass jointly determine the departure from the Einstein coefficient $n-3$. For $\widetilde\alpha>0$, the correction is negative when $\rho(r)<\bar\rho(r)$, vanishes when the two density scales coincide, and is positive when $\rho(r)>\bar\rho(r)$. In particular, a vacuum exterior has $\rho=0$ but $\bar\rho>0$, and therefore gives a coefficient smaller than $n-3$. At a photon orbit, however, $B_\gamma$ is an upper bound on the normalized instability: the exact identity Eq. \ref{eq:shadow-exact} also contains the nonpositive contribution proportional to $\rho_\gamma+p_{t\gamma}$ under the tangential NEC. Combining these observations yields the following proposition.\\

\textit{Proposition (local NEC bound).}
Consider a regular unstable circular null orbit of a static, spherically symmetric EGB solution with $n\geq5$, $\at\geq0$, and $\Xi_\gamma>0$. If the tangential null energy condition $\rhog+\ptg\geq0$ holds at the orbit, then
\begin{equation}
 \eta_\gamma^2\leq B_\gamma
 =n-3+\frac{\rg\Xi'_\gamma}{\Xi_\gamma},
 \label{eq:proposition}
\end{equation}
where $B_\gamma$ is equivalently given by Eq.~\eqref{eq:B-contrast}. Equality is equivalent to tangential NEC saturation. If, in addition, $\rhog\leq\bar\rho_\gamma$, then
\begin{equation}
 \lambda b_\gamma\leq\sqrt{n-3}.
 \label{eq:Einstein-survives}
\end{equation}
For $\at>0$ and $\rhog<\bar\rho_\gamma$,
Eq.~\eqref{eq:Einstein-survives} is strict. The opposite
density ordering raises the upper bound above the
Einstein-gravity value, but does not by itself imply
that the actual instability exceeds that value.
Indeed, combining Eq.~\eqref{eq:shadow-exact} with the
density expression for $B_\gamma$ shows that
$\eta_\gamma^2>n-3$ holds if and only if
\begin{equation}
 \frac{4\at}{(n-2)\rg^2\Xi_\gamma}
 (\rhog-\bar\rho_\gamma)
 >
 \rhog+\ptg.
 \label{eq:violation-threshold}
\end{equation}
Thus, under the tangential NEC, the positive correction
associated with $\rhog>\bar\rho_\gamma$ must exceed
the matter contribution that reduces the instability.
This criterion applies to an orbit of a solution
satisfying the stated assumptions; whether a global
black-hole solution realizes it requires a separate
analysis.

\subsection{Einstein limit and the more restrictive corollary}

For a smooth family with finite limiting orbit data and $\at\to0$, one has $\Xi\to1$, $r\Xi'/\Xi\to0$, $B\to n-3$, and
\begin{equation}
 \eta_\gamma^2\longrightarrow
 n-3-8\pi\rg^2(\rhog+\ptg).
 \label{eq:Einstein-optical-exact}
\end{equation}
This is the exact Einstein identity, including its matter term. Therefore the tangential NEC gives
\begin{equation}
 \lambda b_\gamma\leq\sqrt{n-3},\qquad
 \lambda\leq\sqrt{n-3}\frac{\kg}{\sqrt{\mug}},
 \label{eq:Einstein-limit-bounds}
\end{equation}
consistent with Refs.~\cite{GalloMadler2025,Fairoos2026}. The limit $\widetilde\alpha\to0$ is taken within a smooth family of Einstein-branch solutions at fixed $n\geq5$, with a regular limiting orbit and finite matter and metric data. Then $\Xi_\gamma\to1$, the EGB correction vanishes, and Eq. (\ref{eq:Einstein-optical-exact}) recovers the exact Einstein relation before imposing the energy condition. Eq. (\ref{eq:Einstein-limit-bounds}) follows from the tangential NEC. The same Einstein calculation applies separately in four dimensions, where the coefficient is unity.

It is useful to identify the additional matter condition
under which a bound without an explicit density term
holds. The exact identity~\eqref{eq:lambda-exact} shows
that
$\lambda^2\leq\Omega_\gamma^2F_n(z_\gamma)$
holds if and only if
\begin{equation}
 \Xi_\gamma(\rhog+\ptg)
 \geq\frac{4\at}{(n-2)\rg^2}\rhog.
 \label{eq:matter-condition}
\end{equation}
For an unstable orbit satisfying this condition,
$\lambda^2>0$ implies $F_n(z_\gamma)>0$.
Taking the positive square root therefore gives
\begin{equation}
 \lambda\leq\Omega_\gamma\sqrt{F_n(z_\gamma)}.
 \label{eq:local-bound}
\end{equation}

Since $\Xi_\gamma>0$, the matter condition can
equivalently be written as
\begin{equation}
 \ptg\geq-\rhog+
 \frac{4\at\rhog}{(n-2)\rg^2\Xi_\gamma}.
 \label{eq:pt-condition}
\end{equation}
When $\rhog>0$ and $\at>0$, this condition is strictly
stronger than the tangential NEC, which requires only
$\ptg\geq-\rhog$. In this case, the resulting bound
is also tighter than the principal NEC bound because
$B_\gamma>F_n(z_\gamma)$. For $\rhog=0$ or $\at=0$,
the two matter conditions coincide. If $\rhog<0$
and $\at>0$, the tangential NEC already implies
Eq.~\eqref{eq:matter-condition}, but the principal
bound is then tighter because
$B_\gamma<F_n(z_\gamma)$. Thus
Eq.~\eqref{eq:local-bound} is retained as a
conditional density-free corollary, without imposing
its matter condition in the principal theorem.

To compare this corollary with the Einstein-gravity
value, note that
\begin{equation}
 \frac{\dd F_n}{\dd z}
 =-\frac{2(n-1)}{(1+2z)^3}<0
 \qquad\text{for }1+2z>0.
 \label{eq:Fn-derivative}
\end{equation}
Since $F_n(0)=n-3$, the additional restriction
$z_\gamma\geq0$, together with
Eq.~\eqref{eq:matter-condition}, yields
\begin{equation}
 \lambda b_\gamma
 \leq\sqrt{F_n(z_\gamma)}
 \leq\sqrt{n-3}.
 \label{eq:shadow-bound}
\end{equation}
The restriction $z_\gamma\geq0$ does not follow
from $\Xi_\gamma>0$ alone and is not required for
the principal NEC bound. Alternatively, the
tangential NEC together with
$\rhog\leq\bar\rho_\gamma$ gives the overall
inequality $\lambda b_\gamma\leq\sqrt{n-3}$
directly through Eq.~\eqref{eq:Einstein-survives},
without invoking the density-free corollary.

\section{Vacuum Einstein-branch black holes}
\label{sec:vacuum}

The asymptotically flat vacuum solution on the Einstein branch is \cite{BoulwareDeser1985,Wheeler1986}
%%%%%%%%%%%%%%%%%%%%%%%%%%%%%%%%%%%%%%%%%%%%%%%%%%%%%%%%%%%%%%
\begin{equation}
 \mu(r)=1+\frac{r^2}{2\at}
 \left[1-\sqrt{1+\frac{4\at m}{r^{n-1}}}\right],
 \qquad
 \delta=0,
 \label{eq:BD-metric}
\end{equation}
%%%%%%%%%%%%%%%%%%%%%%%%%%%%%%%%%%%%%%%%%%%%%%%%%%%%%%%%%%%%%%
where $m$ is proportional to the ADM mass,
%%%%%%%%%%%%%%%%%%%%%%%%%%%%%%%%%%%%%%%%%%%%%%%%%%%%%%%%%%%%%%
\begin{equation}
 M_{\rm ADM}=\frac{(n-2)\Omega_{n-2}}{16\pi}\,m.
 \label{eq:ADM-mass}
\end{equation}
%%%%%%%%%%%%%%%%%%%%%%%%%%%%%%%%%%%%%%%%%%%%%%%%%%%%%%%%%%%%%%
The horizon equation is
%%%%%%%%%%%%%%%%%%%%%%%%%%%%%%%%%%%%%%%%%%%%%%%%%%%%%%%%%%%%%%
\begin{equation}
 m=\rH^{n-3}+\at\rH^{n-5}.
 \label{eq:vac-horizon}
\end{equation}
%%%%%%%%%%%%%%%%%%%%%%%%%%%%%%%%%%%%%%%%%%%%%%%%%%%%%%%%%%%%%%
If
%%%%%%%%%%%%%%%%%%%%%%%%%%%%%%%%%%%%%%%%%%%%%%%%%%%%%%%%%%%%%%
\begin{equation}
 S(r)=\sqrt{1+\frac{4\at m}{r^{n-1}}},
 \label{eq:S-vac}
\end{equation}
%%%%%%%%%%%%%%%%%%%%%%%%%%%%%%%%%%%%%%%%%%%%%%%%%%%%%%%%%%%%%%
then differentiation of Eq.~\eqref{eq:BD-metric} gives
%%%%%%%%%%%%%%%%%%%%%%%%%%%%%%%%%%%%%%%%%%%%%%%%%%%%%%%%%%%%%%
\begin{equation}
 \cR(r)=2\mu-r\mu'=2-\frac{(n-1)m}{r^{n-3}S(r)}.
 \label{eq:R-vac}
\end{equation}
%%%%%%%%%%%%%%%%%%%%%%%%%%%%%%%%%%%%%%%%%%%%%%%%%%%%%%%%%%%%%%
Consequently, $\cR(\rg)=0$ reduces to
%%%%%%%%%%%%%%%%%%%%%%%%%%%%%%%%%%%%%%%%%%%%%%%%%%%%%%%%%%%%%%
\begin{equation}
 S_\gamma=\frac{(n-1)m}{2\rg^{n-3}},
 \label{eq:vac-photon}
\end{equation}
%%%%%%%%%%%%%%%%%%%%%%%%%%%%%%%%%%%%%%%%%%%%%%%%%%%%%%%%%%%%%%
or, equivalently,
%%%%%%%%%%%%%%%%%%%%%%%%%%%%%%%%%%%%%%%%%%%%%%%%%%%%%%%%%%%%%%
\begin{equation}
 4\rg^{2n-6}+16\at m\rg^{n-5}=(n-1)^2m^2.
 \label{eq:vac-photon-poly}
\end{equation}
%%%%%%%%%%%%%%%%%%%%%%%%%%%%%%%%%%%%%%%%%%%%%%%%%%%%%%%%%%%%%%
Since $\Xi_\gamma=S_\gamma$, the branch structure is explicit and no division by a vanishing $\Xi_\gamma$ occurs in the regular exterior.

Vacuum has $\mathfrak m=m$, $\rho=p_t=0$ and $S=\Xi$, so $B_\gamma=F_n(z_\gamma)$. The principal NEC bound and its refined corollary are both saturated. For the usual single capture barrier of this family, $b_\gamma=R_{\rm sh}$, and
%%%%%%%%%%%%%%%%%%%%%%%%%%%%%%%%%%%%%%%%%%%%%%%%%%%%%%%%%%%%%%
\begin{equation}
 \lambda=\frac{\sqrt{\mu_\gamma}}{\rg}\sqrt{F_n(z_\gamma)},
 \qquad
 \lambda R_{\rm sh}=\sqrt{F_n(z_\gamma)}.
 \label{eq:vac-saturation}
\end{equation}
%%%%%%%%%%%%%%%%%%%%%%%%%%%%%%%%%%%%%%%%%%%%%%%%%%%%%%%%%%%%%%
The equality shows that $F_n(z_\gamma)$ is not an arbitrary estimate; it is the exact normalized instability of the vacuum orbit.
%%%%%%%%%%%%%%%%%%%%%%%%%%%%%%%%%%%%%%%%%%%%%%%%%%%%%%%%%%%%%%
The Einstein limit also follows directly for arbitrary $n$. As $\at\to0$, $\mu=1-m/r^{n-3}$ and the orbit equation yields
\begin{equation}
 \rg^{n-3}=\frac{n-1}{2}m,\qquad
 \mug=\frac{n-3}{n-1},\qquad
 \eta_\gamma^2=n-3.
 \label{eq:ST-limit}
\end{equation}
Thus the Schwarzschild-Tangherlini solution saturates the Einstein bound, whereas a positive-mass EGB vacuum with $\at>0$ has $\Xi'<0$ and lies strictly below that coefficient. This compares the dimensionless products; a comparison of $\lambda$ between different solutions additionally needs a fixed mass or horizon prescription.

\subsection{Closed Expressions in Five Dimensions}
\label{subsec:5d-vac}

In five dimensions, the horizon relation becomes $m=\rH^2+\at$. The vacuum photon-orbit equation then gives $\rg^4=4m(m-\at)$, yielding
\begin{equation}
\begin{aligned}
\rg^2&=2\rH\sqrt{\rH^2+\at},\\
\mug&=\frac{\rH}{\rH+\sqrt{\rH^2+\at}},\\
\Xi_\gamma&=\sqrt{1+\frac{\at}{\rH^2}}.
\end{aligned}
\label{eq:5d-orbit-data}
\end{equation}
The unique positive photon-orbit radius lies outside the horizon and determines the capture threshold. Hence $R_{\rm sh}^2=\rg^2/\mug$, giving
\begin{equation}
R_{\rm sh}^2
=
2\sqrt{\rH^2+\at}
\left(\rH+\sqrt{\rH^2+\at}\right).
\label{eq:5d-shadow}
\end{equation}
Since $F_5(z_\gamma)=2/\Xi_\gamma^2$ and the NEC bound is saturated in vacuum, the normalized instability is
\begin{equation}
\lambda R_{\rm sh}
=
\sqrt{\frac{2\rH^2}{\rH^2+\at}}.
\label{eq:5d-lamR}
\end{equation}
Combining these expressions yields
\begin{equation}
\lambda^2
=
\frac{\rH^2}
{(\rH^2+\at)^{3/2}
\left(\rH+\sqrt{\rH^2+\at}\right)}.
\label{eq:5d-lambda}
\end{equation}
The Einstein limit $\at\to0$ recovers the five-dimensional Schwarzschild--Tangherlini results,
\begin{equation}
R_{\rm sh}\longrightarrow2\rH,
\qquad
\lambda R_{\rm sh}\longrightarrow\sqrt{2},
\qquad
\lambda\longrightarrow\frac{1}{\sqrt{2}\,\rH}.
\end{equation}

Eq.~\eqref{eq:5d-lamR} shows that increasing the coupling at fixed horizon radius decreases the instability per unit orbital angle. The same conclusion holds at fixed mass parameter, since $\rH^2=m-\at$ gives
\begin{equation}
\lambda R_{\rm sh}
=
\sqrt{2\left(1-\frac{\at}{m}\right)},
\qquad
0\leq\at<m.
\label{eq:5d-fixedm-product}
\end{equation}
Thus, in either comparison, the instability timescale increases relative to the orbital period. At fixed $m$, the endpoint $\at\to m$ sends $\rH$ to zero and does not describe a regular finite-area remnant. These conclusions concern null-geodesic instability; stability and hyperbolicity of gravitational perturbations require separate analysis~\cite{Reall2014,Konoplya2017}.

\section{A matter check: the Maxwell field}
\label{sec:maxwell}

The electrically charged Einstein-branch solution provides a
nonvacuum test of both the principal NEC bound and its refined
corollary \cite{Wiltshire1986,MaLu2020}. With the charge
normalization derived in Appendix~\ref{app:maxwell},
\begin{equation}
 \rho=-p_r=p_t
 =\frac{(n-2)(n-3)q^2}{16\pi r^{2n-4}}.
 \label{eq:Maxwell-stress}
\end{equation}
Thus $\rho+p_t=2\rho$, and the principal NEC bound is strict
for $q\neq0$. Its exact gap is
\begin{equation}
 B_\gamma-\eta_\gamma^2
 =\frac{16\pi\rg^2\rhog}{\Xi_\gamma}
 =\frac{(n-2)(n-3)q^2}
 {\Xi_\gamma\rg^{2n-6}}.
 \label{eq:Maxwell-gap}
\end{equation}
This gap is positive on the regular Einstein branch, where
$\Xi_\gamma>0$. It illustrates why the positive density term
retained in the upper bound does not, by itself, describe the
full effect of matter on the actual instability.

The mass polynomial of this solution is
\begin{equation}
 \mathfrak m(r)=m-\frac{q^2}{r^{n-3}},
 \label{eq:Maxwell-mass}
\end{equation}
where $m$ is the constant asymptotic mass parameter.
Differentiation gives
\begin{equation}
 r\mathfrak m'-(n-1)\mathfrak m
 =\frac{2(n-2)q^2}{r^{n-3}}-(n-1)m.
 \label{eq:Maxwell-contrast}
\end{equation}
Together with the mass equation, these expressions provide
an independent check of the density normalization entering
Eq.~\eqref{eq:B-gradient}.

Substituting Eq.~\eqref{eq:Maxwell-stress} into the matter
bracket in Eq.~\eqref{eq:lambda-exact} yields
\begin{equation}
\begin{aligned}
 &\Xi_\gamma(\rhog+\ptg)
 -\frac{4\at\rhog}{(n-2)\rg^2}\\
 &\qquad
 =2\rhog\left(
 \Xi_\gamma-\frac{2\at}{(n-2)\rg^2}
 \right).
\end{aligned}
 \label{eq:Maxwell-bracket}
\end{equation}
In five dimensions, this bracket is strictly positive for
nonzero charge in the regular nonextremal family, as follows
directly from the exact solution. The nonextremal outer
horizon satisfies $q^2<\rH^4$, and an exterior photon orbit
has $\rg>\rH$. The horizon and photon-orbit equations give
\begin{equation}
 m=\rH^2+\at+\frac{q^2}{\rH^2},
\end{equation}
and
\begin{equation}
\begin{aligned}
 \rg^2\Xi_\gamma
 &=2m-\frac{3q^2}{\rg^2}\\
 &\geq 2\rH^2+2\at-\frac{q^2}{\rH^2}\\
 &>\rH^2+2\at
 >\frac{2\at}{3}.
\end{aligned}
 \label{eq:D-proof}
\end{equation}
The first inequality allows equality when $q=0$; the
remaining inequalities follow from nonextremality,
$\rH>0$, and $\at\geq0$. Consequently,
$\Xi_\gamma-2\at/(3\rg^2)>0$, so the refined corollary
also holds strictly for $q\neq0$.

The density criterion provides another route to the
Einstein optical bound. In five dimensions,
Eq.~\eqref{eq:Maxwell-contrast} gives
\begin{equation}
 \rg^2\left(
 \rg\mathfrak m'_\gamma-4\mathfrak m_\gamma
 \right)
 =6q^2-4m\rg^2<0.
 \label{eq:Maxwell-contrast5}
\end{equation}
Here $\mathfrak m'_\gamma$ denotes the radial derivative
of $\mathfrak m(r)$ evaluated at $r=\rg$. The sign follows
from
\begin{equation}
 m\rg^2>m\rH^2
 =\rH^4+\at\rH^2+q^2>2q^2.
\end{equation}
Hence $\rhog<\bar\rho_\gamma$ at every exterior photon
orbit in this nonextremal five-dimensional family.
For $\at>0$, the density representation of the bound
therefore gives $B_\gamma<2$; at $\at=0$, it gives
$B_\gamma=2$.

The exact normalized instability and the principal NEC
upper bound are
\begin{equation}
 \eta_\gamma^2
 =\frac{2}{\Xi_\gamma^2}
 -\frac{6q^2}{\rg^6\Xi_\gamma^2}
 \left(\rg^2\Xi_\gamma-\frac{2\at}{3}\right),
 \label{eq:Maxwell-direct-main}
\end{equation}
\begin{equation}
 B_\gamma
 =\frac{2}{\Xi_\gamma^2}
 +\frac{4\at q^2}{\rg^6\Xi_\gamma^2},
 \label{eq:Maxwell-B5}
\end{equation}
with the exact difference
\begin{equation}
 B_\gamma-\eta_\gamma^2
 =\frac{6q^2}{\rg^4\Xi_\gamma}>0
 \qquad(q\neq0).
 \label{eq:Maxwell-gap5}
\end{equation}
The first term in Eq.~\eqref{eq:Maxwell-direct-main} is
$F_5(z_\gamma)=2/\Xi_\gamma^2$. Equation~\eqref{eq:D-proof}
therefore establishes both the refined inequality
$\eta_\gamma^2<F_5(z_\gamma)$ and the principal inequality
$\eta_\gamma^2<B_\gamma$ for nonzero charge.

In the Einstein limit, $\at\to0$ and $\Xi_\gamma\to1$,
so these expressions recover the five-dimensional
Einstein--Maxwell result,
\begin{equation}
 B_\gamma=2,
 \qquad
 \eta_\gamma^2=2-\frac{6q^2}{\rg^4}.
\end{equation}
For $q=0$, both inequalities instead reduce to the
vacuum equality
$\eta_\gamma^2=B_\gamma=2/\Xi_\gamma^2$.
The argument assumes a nonextremal outer horizon and a
simple unstable exterior photon orbit; no extremal
limit is required.
\section{Physical interpretation and scope}
\label{sec:discussion}

\subsection{Instability per orbital angle and optical curvature}

Recall that $\eta_\gamma=\lambda/\Omega_\gamma$, introduced in Eq. (\ref{eq:eta-def}), measures the radial instability per unit accumulated orbital angle. We now use this interpretation to express the NEC bound in terms of growth over an orbital period. Along the reference circular orbit, the angular advance from an initial time $t_0$ is $\Delta\phi=\phi(t)-\phi(t_0)=\Omega_\gamma(t-t_0)$. The growing linear radial mode therefore satisfies
\begin{equation}
\Delta r(t)
=
\Delta r(t_0)e^{\lambda(t-t_0)}
=
\Delta r(t_0)e^{\eta_\gamma\Delta\phi}.
\tag{96}
\end{equation}
Here $\Delta\phi$ is counted continuously through successive revolutions and denotes the angular advance along the reference orbit, rather than an angular perturbation. Over one revolution, the growing mode is amplified by $e^{2\pi\eta_\gamma}$. Eq. (\ref{eq:NEC-optical}) bounds this factor by $e^{2\pi\sqrt{B_\gamma}}$.

Equivalently, the result compares the instability timescale $\tau_{\rm inst}=\lambda^{-1}$ with the orbital period $T_{\rm orb}=2\pi/\Omega_\gamma$, both measured using the same asymptotically normalized time. This comparison is standard in the Lyapunov analysis of circular geodesics~\cite{Cardoso2009}. The NEC bound gives
\begin{equation}
\tau_{\rm inst}
\geq
\frac{1}{\Omega_\gamma\sqrt{B_\gamma}},
\qquad
\frac{\tau_{\rm inst}}{T_{\rm orb}}
=
\frac{1}{2\pi\eta_\gamma}
\geq
\frac{1}{2\pi\sqrt{B_\gamma}}.
\tag{97}
\end{equation}
These relations describe the growing radial mode within the linear neighborhood of the circular orbit. The total time spent near the orbit by an individual ray also depends on its initial displacement and on the extent of the neighborhood being considered; it is therefore not determined by $\lambda^{-1}$ alone.

The same bound has a curvature interpretation. In any orbital plane, null propagation induces the Riemannian optical metric \cite{QiaoLi2022,GalloMadler2025}
\begin{equation}
 \dd\ell_{\rm opt}^2=\frac{\dd r^2}{A\mu}
 +\frac{r^2}{A}\dd\phi^2
 =\dd x^2+b(x)^2\dd\phi^2,
 \label{eq:optical-metric}
\end{equation}
where $\dd x/\dd r=(A\mu)^{-1/2}$ and $b=r/\sqrt A$. The spatial null trajectories are optical geodesics, and $\dd\ell_{\rm opt}=\dd t$ along a future-directed ray. At a circular orbit, $\dd b/\dd x=0$. The Gaussian curvature $\mathcal K=-b^{-1}\dd^2b/\dd x^2$ consequently becomes
\begin{equation}
 \mathcal K_\gamma
 =\frac{\mug}{2}\left(A''_\gamma-\frac{2A_\gamma}{\rg^2}\right)
 =-\lambda^2.
 \label{eq:optical-curvature}
\end{equation}
This relation is geometrical and does not use the EGB equations. The optical Jacobi equation for a transverse separation $y$ is $\dd^2y/\dd t^2+\mathcal K_\gamma y=0$; negative curvature therefore gives exponential radial separation \cite{GalloMadler2025}. The NEC bound is equivalent to
\begin{equation}
 -\Omega_\gamma^2B_\gamma\leq\mathcal K_\gamma<0.
 \label{eq:curvature-bound}
\end{equation}
Given the local geometry and density, the optical curvature cannot be arbitrarily negative. This is a local lower bound, not a universal curvature constant shared by all matter distributions.

\subsection{Tangential null convergence and tidal separation}

A complementary spacetime description distinguishes the radial tidal instability from the total focusing of a bundle. Adopt the Riemann convention $[\nabla_\mu,\nabla_\nu]v^\rho=R^\rho{}_{\sigma\mu\nu}v^\sigma$. At the orbit let $u=A^{-1/2}\partial_t$, $e_{\hat r}=\sqrt\mu\,\partial_r$ and $e_{\hat\phi}=r^{-1}\partial_\phi$. Normalize the tangent null vector by
\begin{equation}
 k=u+e_{\hat\phi},\qquad -u\cdot k=1.
 \label{eq:tangent-null}
\end{equation}
Since $A$ is constant along the circular orbit, this normalization is compatible with an affine parametrization there. It gives $T_{\mu\nu}k^\mu k^\nu=\rho+p_t$.

The screen orthogonal to the null direction has one radial direction and $n-3$ angular directions normal to the orbital plane. Define the tidal matrix $\mathcal E_{ab}=R_{\mu\alpha\nu\beta}e_a^\mu k^\alpha e_b^\nu k^\beta$, so that screen geodesic deviation is $\dd^2\xi^a/\dd s^2=-\mathcal E^a{}_b\xi^b$. At the orbit its eigenvalues are
\begin{equation}
 \mathcal E_{\hat r\hat r}=-\frac{\lambda^2}{A_\gamma},\qquad
 \mathcal E_{\hat a\hat a}=\frac{1}{\rg^2}
 \quad(\hat a=1,\ldots,n-3).
 \label{eq:tidal-eigenvalues}
\end{equation}
Appendix~\ref{app:interpretation} derives these expressions directly. The radial screen basis is parallel transported modulo a multiple of $k$, which does not affect screen components. Taking the screen trace yields the theory-independent identity
\begin{equation}
 R_{\mu\nu}k^\mu k^\nu\big|_\gamma
 =\frac{n-3}{\rg^2}-\frac{\lambda^2}{A_\gamma},
 \label{eq:Ricci-orbit}
\end{equation}
or $\eta_\gamma^2=n-3-\rg^2R_{\mu\nu}k^\mu k^\nu|_\gamma$. Radial exponential separation is therefore compatible with nonnegative total null convergence: positive angular tidal eigenvalues can compensate the negative radial one. The factor $n-3$ counts those angular screen directions.

Combining this geometrical identity with Eqs.~\eqref{eq:shadow-exact} and \eqref{eq:B-gradient} gives an exact EGB relation at the orbit,
\begin{equation}
 R_{\mu\nu}k^\mu k^\nu\big|_\gamma
 =\frac{8\pi(\rhog+\ptg)}{\Xi_\gamma}
 -\frac{\Xi'_\gamma}{\rg\Xi_\gamma}.
 \label{eq:EGB-convergence}
\end{equation}
Thus ordinary tangential NEC entails
\begin{equation}
 R_{\mu\nu}k^\mu k^\nu\big|_\gamma
 \geq-\frac{\Xi'_\gamma}{\rg\Xi_\gamma}.
 \label{eq:convergence-bound}
\end{equation}
Eq.~\eqref{eq:NEC-bound} is precisely this lower bound on tangential null convergence translated into an upper bound on radial instability. When $\Xi'_\gamma\leq0$, the NEC guarantees ordinary null convergence and hence the Einstein optical coefficient. When $\Xi'_\gamma>0$, it permits a negative convergence term but limits its magnitude. Negative convergence is not compulsory: the positive first term in Eq.~\eqref{eq:EGB-convergence} can still dominate.

For an affinely parametrized null congruence, the Raychaudhuri equation reads \cite{Raychaudhuri1955,KontouSanders2020}
\begin{equation}
 \frac{\dd\theta}{\dd s}
 =-\frac{\theta^2}{n-2}-\sigma_{ab}\sigma^{ab}
 +\omega_{ab}\omega^{ab}-R_{\mu\nu}k^\mu k^\nu.
 \label{eq:Raychaudhuri}
\end{equation}
Here $\theta$, $\sigma_{ab}$ and $\omega_{ab}$ are the expansion, shear and twist. The EGB equations modify the relation between matter and the last curvature term, not this geometrical evolution equation. The bound concerns one tidal eigenvalue through the constrained trace; it does not identify $\lambda$ with $\theta$ or remove the shear contribution.

This tangential statement must be distinguished from radial focusing. Nozawa and Maeda showed that, in spherical EGB gravity, NEC implies null convergence for radial null vectors on the GR branch \cite{NozawaMaeda2008}. Indeed, for $\ell=u+e_{\hat r}$ the static equations give
\begin{equation}
 R_{\mu\nu}\ell^\mu\ell^\nu
 =-\frac{(n-2)\mu\delta'}{r}
 =\frac{8\pi(\rho+p_r)}{\Xi}.
 \label{eq:radial-convergence}
\end{equation}
The photon-orbit tangent is angular, and its contraction instead contains the derivative term in Eq.~\eqref{eq:EGB-convergence}. There is no conflict between the radial focusing theorem and our density-dependent tangential bound.

\subsection{Meaning of the density contrast and the Lovelock response}

The two terms in Eq.~\eqref{eq:B-def} should be interpreted together. The departure of $F_n$ relative to $n-3$ encodes the enclosed mass through $z(1+z)$, while the explicit density term encodes the local mass gradient. Their combined correction is proportional to $\rho-\bar\rho$. Equivalently,
\begin{equation}
 \bar\rho'=\frac{n-1}{r}(\rho-\bar\rho).
 \label{eq:mean-density-gradient}
\end{equation}
The sign of the EGB shift in the NEC envelope therefore records whether the enclosed-mass density scale is rising or falling at the photon sphere. Vacuum has $\rho=0$ and positive $\bar\rho$, so the envelope is reduced. An exterior shell can change that sign; positivity of density alone cannot decide it. None of these comparisons implies monotonicity of $\lambda$ between distinct solutions at fixed mass, because $r_\gamma$, $A_\gamma$ and the matter variables also change.

To connect this observation to the EGB equations themselves, set $\psi=X/r^2$ and $W(\psi)=\psi+\at\psi^2$. Then $W=\mathfrak m/r^{n-1}$ and $W_{,\psi}=\Xi$. At fixed $r$, this derivative is the response of the spherical mass polynomial to a change in curvature. Define
\begin{equation}
 \mathcal T(r)=r^{n-3}\Xi(r)
 =\left.\frac{\partial\mathfrak m}{\partial X}\right|_r.
 \label{eq:T-def}
\end{equation}
The NEC bound can be written compactly as
\begin{equation}
 \eta_\gamma^2\leq
 \frac{\rg\mathcal T'_\gamma}{\mathcal T_\gamma}.
 \label{eq:T-bound}
\end{equation}
The EGB correction is thus the change in the logarithmic radial growth of this mass-polynomial response from its Einstein value $n-3$. This description refers to the spherical background equations and does not posit a single effective Newton constant for all perturbations.

The function $\mathcal T=r^{n-3}W_{,\psi}$ also appears in the master equations for static vacuum Lovelock perturbations \cite{TakahashiSoda2010}. In that setting the vector-sector angular characteristic coefficient is
\begin{equation}
 c_V(r)=\frac{r\mathcal T'}{(n-3)\mathcal T}.
 \label{eq:vector-characteristic}
\end{equation}
Vacuum saturation implies $\eta_\gamma^2=(n-3)c_V(\rg)$. This links two quantities built from the same background response, but it does not equate photon and graviton circular orbits. The latter involve sector-dependent effective metrics \cite{Reall2014}. We do not extend this vacuum perturbation formula to arbitrary anisotropic matter, nor infer hyperbolicity of every gravitational sector from $B_\gamma>0$ at one radius.

\subsection{Lensing, test-field damping, and limitations}

For a nondegenerate photon sphere that controls the strong-deflection scattering branch, the deflection angle has the form
\begin{equation}
 \Delta\phi(b)=-\bar a\ln(b/b_\gamma-1)+\bar b+o(1).
 \label{eq:strong-deflection}
\end{equation}
The usual orbital-plane calculation gives $\bar a=\Omega_\gamma/\lambda=1/\eta_\gamma$ \cite{Bozza2002,Stefanov2010}. The same planar geodesic calculation applies in spherical higher-dimensional geometry; the interpretation of measured image fluxes would require the appropriate observer geometry. The NEC bound gives
\begin{equation}
 \bar a\geq B_\gamma^{-1/2}.
 \label{eq:lensing-bound}
\end{equation}
It therefore limits the coefficient governing the logarithmic accumulation of deflection near the critical impact parameter, without imposing the refined matter condition.

For photon-sphere modes of a minimally coupled test field, when the eikonal geodesic correspondence applies, one has \cite{Cardoso2009}
\begin{equation}
 \omega_{\ell p}\simeq\ell\Omega_\gamma
 -\mathrm{i}(p+\tfrac12)\lambda.
 \label{eq:eikonal}
\end{equation}
Here $p$ labels the overtone and the expression retains the leading large-$\ell$ terms. It follows that
\begin{equation}
 \frac{|\operatorname{Im}\omega_{\ell p}|}
 {(p+\tfrac12)\Omega_\gamma}\lesssim\sqrt{B_\gamma}.
 \label{eq:damping-bound}
\end{equation}
Subleading corrections and modes not associated with the relevant potential maximum are not bounded by this argument. Gravitational EGB perturbations require their own characteristic metrics and can have eikonal instabilities \cite{DottiGleiser2005,TakahashiSoda2010,Reall2014,KonoplyaStuchlik2017,Konoplya2017}. The background light-ray bound is not a universal bound on gravitational-mode damping or a thermal quantum-chaos inequality.

Finally, the principal result is local. It does not bound density itself or give a universal number for all EGB matter configurations. A bound using only ADM mass, horizon radius or coupling requires further global information. All divisions by $\Xi$ require the stated regular branch, and neither the NEC theorem nor its optical interpretation extends across $\Xi=0$ by continuity alone. The density-contrast corollary and the more restrictive condition~\eqref{eq:matter-condition} are useful additional results, not assumptions hidden in the principal NEC bound.

\section{Conclusions}
\label{sec:conclusion}

We have derived a local upper bound on the coordinate-time Lyapunov exponent of unstable circular null orbits in static, spherically symmetric, asymptotically flat EGB gravity with $n\geq5$ and $\at\geq0$. Within the nondegenerate Einstein branch, where $\Xi_\gamma>0$, the only matter inequality required for the principal bound is the tangential null energy condition, $\rhog+\ptg\geq0$. Neither nonnegative density nor an additional density--pressure inequality is required. The result can be written as
\begin{equation}
 \frac{\lambda}{\Omega_\gamma}\leq\sqrt{B_\gamma},
\end{equation}
where $\Omega_\gamma$ is the orbital angular frequency and $B_\gamma$ is the dimensionless coefficient defined in Eq.~\eqref{eq:B-def}. The ratio on the left measures radial instability per unit orbital angle. The exact identity in Eq.~\eqref{eq:shadow-exact} shows that equality holds precisely when the tangential NEC is saturated at the orbit. Vacuum Einstein-branch black holes provide saturating solutions, while the Maxwell field produces a strictly positive gap for nonzero charge.

The density contribution retained in the bound has a direct interpretation through the spherical mass relation. The coefficient admits the equivalent forms
\begin{equation}
\begin{aligned}
 B_\gamma
 &=n-3+\frac{\rg\Xi'_\gamma}{\Xi_\gamma}\\
 &=n-3+\frac{32\pi\at}{(n-2)\Xi_\gamma^2}
 \left(\rhog-\bar\rho_\gamma\right),
\end{aligned}
\end{equation}
where $\Xi$ is the dimensionless geometrical function defined in Eq.~\eqref{eq:XD-def}, and $\bar\rho_\gamma$ is the enclosed generalized Misner--Sharp mass divided by the areal volume. This enclosed mass includes the horizon contribution. The departure of the bound from the Einstein-gravity value $n-3$ is therefore determined by the local density relative to this enclosed mass density scale. Under the tangential NEC, $\rhog\leq\bar\rho_\gamma$ is sufficient to preserve $\lambda/\Omega_\gamma\leq\sqrt{n-3}$ at nonzero coupling. A positive density difference alone does not imply a violation, because the exact instability also contains the nonpositive contribution proportional to $\rhog+\ptg$.

The geometrical interpretations clarify what is being bounded. At an unstable circular null orbit, the Gaussian curvature of the optical orbital plane equals $-\lambda^2$, so the instability bound limits how negative this curvature can be. In the spacetime description, radial tidal separation contributes negatively to tangential null convergence, while the $n-3$ transverse angular directions contribute positively. Radial instability is therefore compatible with nonnegative total null convergence. The EGB field equations modify the relation between this convergence and the matter NEC through the radial variation of $\Xi$, as displayed in Eq.~\eqref{eq:EGB-convergence}.

For a smooth Einstein limit with finite limiting local data, the exact identity reduces to its Einstein-gravity form and the bound becomes $\lambda/\Omega_\gamma\leq\sqrt{n-3}$. The vacuum and charged solutions provide explicit checks of the geometrical coefficients and matter normalization. The density-free bound obtained under Eq.~\eqref{eq:matter-condition} remains a conditional corollary; its additional matter restriction is not part of the principal result.

These statements concern local null-geodesic instability. When the selected orbit controls capture from infinity, $\Omega_\gamma^{-1}=R_{\rm sh}$ and the result becomes a bound on $\lambda R_{\rm sh}$. Its consequences for strong-deflection lensing and test-field eikonal damping apply under the corresponding assumptions described above. The bound does not by itself establish stability or hyperbolicity of gravitational perturbations, or a universal estimate determined solely by the horizon radius or asymptotic mass. Further work may identify global matter solutions that realize positive density differences at capture-controlling orbits and determine whether their actual instability can exceed the Einstein-gravity value.

\begin{acknowledgments}
The work of CF is supported by the Institutional Faculty Research Seed Grant Scheme (IFRSG-2026) of the TKM College Trust. CF and CLA gratefully acknowledge the course ``Advanced Topics in General Relativity for Multimessenger Observations'' held at IUCAA, Pune, in October 2026. The lectures and discussions during the course substantially enriched their understanding and contributed to significant improvements in this work.
\end{acknowledgments}

\appendix

\section{Checks of the Reduced EGB Field Equations}
\label{app:reduced-equations}

For completeness, this appendix shows how the two reduced equations used in the main text can be checked without recomputing the full Lovelock tensor.  Substitution of Eq.~\eqref{eq:metric} into Eq.~\eqref{eq:EGB-field} gives Eq.~\eqref{eq:rho-equation} and Eq.~\eqref{eq:pr-unrearranged}; equivalent component equations are standard in the EGB literature \cite{Wheeler1986,Wiltshire1986,GalloVillanueva2015}.  The generalized mass polynomial provides an independent check of the temporal equation, while the sum of the temporal and radial equations verifies the equation for the redshift function \cite{MaedaNozawa2008}.
\begin{equation}
 \mathfrak m=r^{n-3}X+\at r^{n-5}X^2.
 \label{eq:mass-app}
\end{equation}
Its derivative is
\begin{align}
 \mathfrak m'={}&r^{n-6}\bigl\{
 r^2[rX'+(n-3)X]
 \nonumber\\
 &\hspace{18mm}+\at X[2rX'+(n-5)X]\bigr\}.
 \label{eq:mass-app-derivative}
\end{align}
The right-hand side is exactly $16\pi r^{n-2}\rho/(n-2)$ when Eq.~\eqref{eq:rho-equation} is used. This reproduces Eq.~\eqref{eq:mass-derivative} and checks the coefficients of the $X'$ and $X^2$ terms in the temporal equation.

The lapse equation provides a second check.  Adding the displayed temporal and radial equations cancels the common $X$ and $X'$ terms and leaves
\begin{equation}
 8\pi(\rho+p_r)
 =-\frac{(n-2)\mu\delta'}{r^3}
 \left(r^2+2\at X\right).
 \label{eq:lapse-app-intermediate}
\end{equation}
Solving for $\delta'$ gives Eq.~\eqref{eq:delta-equation}.  Finally, the radial conservation equation follows directly from the $\nu=r$ component of $\nabla_\mu T^\mu{}_\nu=0$:
\begin{equation}
 p_r'+\frac{A'}{2A}(\rho+p_r)
 +\frac{n-2}{r}(p_r-p_t)=0,
 \label{eq:conservation-app}
\end{equation}
which is Eq.~\eqref{eq:conservation}.  Taken together, these relations provide complementary checks of the normalization used in the main calculation. 

\section{Algebra Leading to the Exact Identity}
\label{app:algebra}

This appendix gives the intermediate algebra leading from the
radial field equation to Eq.~\eqref{eq:Rprime-master}, with the
coefficients and signs displayed explicitly. Throughout, we use
$\Xi=1+2\at X/r^2$ and $X=1-\mu$, as defined in
Eq.~\eqref{eq:XD-def}.

Multiplying Eq.~\eqref{eq:pr-R} by $2r^4$ and collecting
the terms proportional to $\cR$ gives $\Delta\cR=Q$, where
\begin{equation}
 \cR=\frac{Q}{\Delta},
 \qquad
 \Delta=2\at(\mu-1)-r^2=-r^2\Xi,
 \label{eq:R-Q}
\end{equation}
and
\begin{equation}
\begin{aligned}
 Q={}&(n-1)\at\mu^2\\
 &-\bigl[(n-1)r^2+2(n-3)\at\bigr]\mu\\
 &+(n-5)\at+(n-3)r^2
 +\frac{16\pi r^4}{n-2}p_r.
\end{aligned}
 \label{eq:Q-def}
\end{equation}
At the photon sphere, $\cR_\gamma=0$. Since
$\Delta_\gamma=-\rg^2\Xi_\gamma\neq0$ on the regular
Einstein branch, it follows that $Q_\gamma=0$.
Differentiating Eq.~\eqref{eq:R-Q} before imposing the
circular-orbit condition gives
\begin{equation}
 \cR'
 =\frac{Q'\Delta-Q\Delta'}{\Delta^2}.
\end{equation}
The term proportional to $Q$ therefore vanishes at the
photon sphere, leaving
\begin{equation}
 \cR'_\gamma
 =-\frac{Q'_\gamma}{\rg^2\Xi_\gamma}.
 \label{eq:Rprime-Q}
\end{equation}

The temporal field equation,
Eq.~\eqref{eq:rho-equation}, determines $\mu'_\gamma$:
\begin{equation}
 \mu'_\gamma
 =
 \frac{(n-3)\rg^2X_\gamma+(n-5)\at X_\gamma^2
 -\dfrac{16\pi\rhog\rg^4}{n-2}}
 {\rg^3\Xi_\gamma}.
 \label{eq:mup-app}
\end{equation}
Meanwhile, solving $Q_\gamma=0$ for the radial pressure
gives
\begin{equation}
\begin{aligned}
 \prg={}&\frac{n-2}{16\pi\rg^4}
 \bigl[2\rg^2+4\at X_\gamma\\
 &\qquad -(n-1)\rg^2X_\gamma
 -(n-1)\at X_\gamma^2\bigr],
\end{aligned}
 \label{eq:pr-app}
\end{equation}
where $X_\gamma=1-\mu_\gamma$.

Differentiating Eq.~\eqref{eq:Q-def} and using
Eq.~\eqref{eq:conservation-gamma} to eliminate
$p'_{r\gamma}$ yields
\begin{equation}
\begin{aligned}
 Q'_\gamma={}&
 \bigl[2(n-1)\at\mug-(n-1)\rg^2\\
 &\qquad -2(n-3)\at\bigr]\mu'_\gamma\\
 &-2(n-1)\rg\mug+2(n-3)\rg\\
 &+\frac{16\pi\rg^3}{n-2}
 \bigl[-\rhog+(n-2)\ptg-(n-5)\prg\bigr].
\end{aligned}
 \label{eq:Qp-app}
\end{equation}
In particular, the coefficient of $\prg$ in the last line
comes from combining the derivative of $r^4p_r$ with
the $-(n-1)\prg$ term in the conservation equation.

To track the density terms explicitly, first rewrite the
coefficient multiplying $\mu'_\gamma$ as
\begin{equation}
\begin{aligned}
 &2(n-1)\at\mug-(n-1)\rg^2-2(n-3)\at\\
 &\qquad=4\at-(n-1)\rg^2\Xi_\gamma.
\end{aligned}
\end{equation}
After substituting Eq.~\eqref{eq:mup-app} and multiplying
Eq.~\eqref{eq:Qp-app} by $\rg^3\Xi_\gamma$, the density
contribution from $\mu'_\gamma$ combines with the explicit
$-\rhog$ term according to
\begin{equation}
\begin{aligned}
 &-\frac{16\pi\rg^4\rhog}{n-2}
 \bigl[4\at-(n-1)\rg^2\Xi_\gamma\bigr]
 -\frac{16\pi\rg^6\Xi_\gamma\rhog}{n-2}\\
 &\qquad
 =16\pi\rg^6\Xi_\gamma\rhog
 -\frac{64\pi\at\rg^4}{n-2}\rhog.
\end{aligned}
\end{equation}
The tangential-pressure term contributes
$16\pi\rg^6\Xi_\gamma\ptg$. Substituting
Eq.~\eqref{eq:pr-app} into the remaining terms and
collecting the geometrical contributions therefore gives
\begin{equation}
\begin{aligned}
 \rg^3\Xi_\gamma Q'_\gamma={}&
 -2(n-3)\rg^4\\
 &-4(n-5)\at\rg^2X_\gamma\\
 &-4(n-5)\at^2X_\gamma^2\\
 &+16\pi\rg^6\Xi_\gamma(\rhog+\ptg)\\
 &-\frac{64\pi\at\rg^4}{n-2}\rhog.
\end{aligned}
 \label{eq:Qp-common-numerator}
\end{equation}

Writing
$Q'_\gamma=Q'_{\gamma,{\rm geom}}+
Q'_{\gamma,{\rm mat}}$, we can separate this result into
\begin{equation}
\begin{aligned}
 Q'_{\gamma,{\rm geom}}
 ={}&-\frac{2}{\rg^3\Xi_\gamma}
 \bigl[(n-3)\rg^4\\
 &\qquad+2(n-5)\at\rg^2X_\gamma\\
 &\qquad+2(n-5)\at^2X_\gamma^2\bigr],
\end{aligned}
 \label{eq:Qp-geom}
\end{equation}
and
\begin{equation}
 Q'_{\gamma,{\rm mat}}
 =
 \frac{16\pi\rg^3}{\Xi_\gamma}
 \left[
 \Xi_\gamma(\rhog+\ptg)
 -\frac{4\at}{(n-2)\rg^2}\rhog
 \right].
 \label{eq:Qp-mat}
\end{equation}

It remains to express the geometrical contribution in
terms of $z_\gamma$. Using
$\at X_\gamma=z_\gamma\rg^2$ and
$\Xi_\gamma=1+2z_\gamma$, Eq.~\eqref{eq:Qp-geom} gives
\begin{equation}
\begin{aligned}
 -\frac{Q'_{\gamma,{\rm geom}}}{\rg^2\Xi_\gamma}
 &=
 \frac{2}{\rg}
 \frac{(n-3)+2(n-5)z_\gamma
 +2(n-5)z_\gamma^2}
 {(1+2z_\gamma)^2}\\
 &=\frac{2}{\rg}F_n(z_\gamma),
\end{aligned}
\end{equation}
where the second equality uses Eq.~\eqref{eq:Fn}.
Similarly, Eq.~\eqref{eq:Qp-mat} yields
\begin{equation}
\begin{aligned}
 -\frac{Q'_{\gamma,{\rm mat}}}{\rg^2\Xi_\gamma}
 ={}&-\frac{16\pi\rg}{\Xi_\gamma^2}
 \left[
 \Xi_\gamma(\rhog+\ptg)
 -\frac{4\at}{(n-2)\rg^2}\rhog
 \right].
\end{aligned}
\end{equation}
Adding these contributions and using
Eq.~\eqref{eq:Rprime-Q} gives
\begin{equation}
\begin{aligned}
 \cR'_\gamma={}&\frac{2}{\rg}F_n(z_\gamma)\\
 &-\frac{16\pi\rg}{\Xi_\gamma^2}
 \left[
 \Xi_\gamma(\rhog+\ptg)
 -\frac{4\at}{(n-2)\rg^2}\rhog
 \right],
\end{aligned}
\end{equation}
which reproduces Eq.~\eqref{eq:Rprime-master}.
No energy condition has been imposed in this derivation.

As an immediate check, $\at\to0$ gives $\Xi_\gamma\to1$
and $F_n(z_\gamma)\to n-3$, recovering
\begin{equation}
 \cR'_\gamma
 \longrightarrow
 \frac{2(n-3)}{\rg}
 -16\pi\rg(\rhog+\ptg),
\end{equation}
in agreement with Eq.~\eqref{eq:Einstein-Rprime}.
The explicit-solution checks in Appendix~\ref{app:checks}
provide complementary verification of the geometrical
coefficients and the matter normalization.

\section{Maxwell Normalization and the Five-dimensional Consistency Proof}
\label{app:maxwell}

This appendix gives the Maxwell normalization and the details of the charged example discussed in Sec.~\ref{sec:maxwell}. Let $A_\mu$ denote the electromagnetic potential, distinct from the metric function $A(r)$, with field strength
\begin{equation}
F_{\mu\nu}
=
\nabla_\mu A_\nu-\nabla_\nu A_\mu.
\end{equation}
The Maxwell action is
\begin{equation}
S_{\rm m}
=
-\frac{1}{16\pi}
\int \dd^n x\sqrt{-g}\,
F_{\mu\nu}F^{\mu\nu}.
\label{eq:Maxwell-action}
\end{equation}
Variation with respect to $A_\mu$ gives $\nabla_\mu F^{\mu\nu}=0$, while variation with respect to the metric gives
\begin{equation}
T_{\mu\nu}
=
\frac{1}{4\pi}
\left(
F_{\mu\rho}F_\nu{}^\rho
-\frac14 g_{\mu\nu}F_{\rho\sigma}F^{\rho\sigma}
\right).
\label{eq:Maxwell-tensor}
\end{equation}

For a purely radial electric field, the stress tensor satisfies $\rho+p_r=0$. The reduced EGB equations therefore imply $\delta'=0$ in the regular static exterior. Asymptotic normalization fixes this constant to $\delta=0$. The Maxwell equation then reduces to
\begin{equation}
\partial_r\left(r^{n-2}F^{rt}\right)=0.
\label{eq:Maxwell-radial-equation}
\end{equation}
Since $g^{rr}=\mu$, $g^{tt}=-1/\mu$, and $F_{rt}=-F_{tr}$, one has $F^{rt}=F_{tr}$. Defining the integration constant in terms of the charge parameter $q$ gives
\begin{equation}
F_{tr}
=
\sqrt{\frac{(n-2)(n-3)}{2}}\,
\frac{q}{r^{n-2}}.
\label{eq:Maxwell-field}
\end{equation}
With the corresponding flux convention, the conserved electric charge is
\begin{equation}
Q_{\rm e}
=
\frac{\Omega_{n-2}}{4\pi}r^{n-2}F^{rt}
=
\frac{\Omega_{n-2}}{4\pi}
\sqrt{\frac{(n-2)(n-3)}{2}}\,q.
\label{eq:physical-charge}
\end{equation}
Substituting Eq.~\eqref{eq:Maxwell-field} into the stress tensor yields
\begin{equation}
\rho=-p_r=p_t
=
\frac{(n-2)(n-3)q^2}{16\pi r^{2n-4}},
\end{equation}
which reproduces Eq.~\eqref{eq:Maxwell-stress} and fixes the normalization of the matter terms.

The asymptotically flat Einstein-branch solution is
\cite{Wiltshire1986,MaLu2020}
\begin{equation}
\mu(r)
=
1+\frac{r^2}{2\at}
\left[
1-\sqrt{
1+4\at\left(
\frac{m}{r^{n-1}}-\frac{q^2}{r^{2n-4}}
\right)}
\right],
\label{eq:charged-metric}
\end{equation}
where $m$ is the asymptotic mass parameter and the value at $\at=0$ is understood as the smooth limit. Its mass polynomial is $\mathfrak m(r)=m-q^2/r^{n-3}$, whose derivative also reproduces the density through Eq.~\eqref{eq:mass-derivative}.

We now specialize to five dimensions. The square root appearing in the metric is precisely the previously defined function $\Xi$:
\begin{equation}
\Xi(r)
=
\sqrt{
1+4\at\left(
\frac{m}{r^4}-\frac{q^2}{r^6}
\right)}.
\label{eq:S-charged}
\end{equation}
At the horizon, $\mu(\rH)=0$ gives $\Xi(\rH)=1+2\at/\rH^2$, and hence
\begin{equation}
m
=
\rH^2+\at+\frac{q^2}{\rH^2}.
\label{eq:charged-horizon}
\end{equation}
Differentiating the metric and using this relation gives
\begin{equation}
\mu'(\rH)
=
\frac{2(\rH^4-q^2)}
{\rH^3(\rH^2+2\at)}.
\label{eq:charged-horizon-derivative}
\end{equation}
For $\rH>0$ and $\at\geq0$, the nonextremal outer-horizon condition is therefore $q^2<\rH^4$.

Direct differentiation also simplifies the photon-sphere characteristic function to
\begin{equation}
\cR(r)
=
2\mu-r\mu'
=
2-\frac{4m/r^2-6q^2/r^4}{\Xi(r)}.
\label{eq:R-charged-intermediate}
\end{equation}
This expression remains regular in the Einstein limit. The circular-orbit condition $\cR(\rg)=0$ becomes
\begin{equation}
\Xi_\gamma
=
\frac{2m}{\rg^2}
-\frac{3q^2}{\rg^4}.
\label{eq:charged-photon-S}
\end{equation}
Squaring this relation and using Eq.~\eqref{eq:S-charged} gives
\begin{equation}
\rg^8
+4m(\at-m)\rg^4
+4(3m-\at)q^2\rg^2
-9q^4
=
0.
\label{eq:charged-quartic}
\end{equation}
An admissible exterior root must satisfy $\rg>\rH$ and $2m\rg^2-3q^2>0$. The latter condition restores the sign information lost on squaring the orbit equation.

An exterior circular orbit exists because
$\cR(\rH)=-\rH\mu'(\rH)<0$, whereas $\cR(r)\to2$ at infinity. The local instability calculation applies to any simple unstable exterior root. For the shadow interpretation, the selected orbit must determine the capture threshold, corresponding to a global maximum of $\mu/r^2$. We assume that this maximum is nondegenerate; then $\cR'_\gamma>0$ and Eq.~\eqref{eq:lambda-R} gives $\lambda^2>0$. This choice also ensures $b_\gamma=R_{\rm sh}$.

The orbit equation gives the useful relation
\begin{equation}
\rg^2\Xi_\gamma
=
2m-\frac{3q^2}{\rg^2}.
\label{eq:D-charged}
\end{equation}
Using $\rg>\rH$, Eq.~\eqref{eq:charged-horizon}, and $q^2<\rH^4$, we obtain
\begin{equation}
\begin{aligned}
\rg^2\Xi_\gamma
&\geq
2\rH^2+2\at-\frac{q^2}{\rH^2}\\
&>
\rH^2+2\at
>
\frac{2\at}{3}.
\end{aligned}
\label{eq:D-proof-app}
\end{equation}
The first inequality is strict for $q\neq0$ and becomes an equality at $q=0$. The remaining inequalities are strict throughout the stated nonextremal parameter range.

In five dimensions, $\rho_\gamma+p_{t\gamma}=2\rho_\gamma$ and $F_5(z_\gamma)=2/\Xi_\gamma^2$. Substitution into the exact instability identity gives
\begin{equation}
\eta_\gamma^2
=
\frac{2}{\Xi_\gamma^2}
-
\frac{6q^2}{\rg^6\Xi_\gamma^2}
\left(
\rg^2\Xi_\gamma-\frac{2\at}{3}
\right).
\label{eq:charged-lamR}
\end{equation}
For the capture orbit, the left-hand side is $(\lambda R_{\rm sh})^2$. Equation~\eqref{eq:D-proof-app} shows that the subtracted term is positive whenever $q\neq0$. Thus the tighter conditional bound is satisfied strictly in this nonextremal charged family.

The principal NEC bound has the coefficient
\begin{equation}
B_\gamma
=
\frac{2}{\Xi_\gamma^2}
+
\frac{4\at q^2}{\rg^6\Xi_\gamma^2},
\end{equation}
and its exact gap is
\begin{equation}
B_\gamma-\eta_\gamma^2
=
\frac{6q^2}{\rg^4\Xi_\gamma}
>
0
\qquad(q\neq0).
\end{equation}
The gap vanishes in the vacuum limit. In the Einstein limit, $\Xi_\gamma\to1$ and Eq.~\eqref{eq:charged-lamR} reduces to
\begin{equation}
\eta_\gamma^2
=
2-\frac{6q^2}{\rg^4},
\qquad
B_\gamma=2,
\end{equation}
recovering the five-dimensional Einstein--Maxwell result.
\section{Independent Checks}
\label{app:checks}

We perform three checks of the exact identity derived in Appendix~\ref{app:algebra}. First, setting $\at=0$ recovers Eq.~\eqref{eq:Einstein-Rprime}, including the coefficient of the matter term.

Second, the explicit solutions allow the Lyapunov exponent to be calculated directly from the metric. For $\delta=0$, the characteristic function is $\cR=2\mu-r\mu'$, so
\begin{equation}
\cR'_\gamma
=
\mu'_\gamma-\rg\mu''_\gamma
=
\frac{2\mug}{\rg}-\rg\mu''_\gamma,
\label{eq:Rprime-direct}
\end{equation}
where the circular-orbit condition $\rg\mu'_\gamma=2\mug$ was used. Substituting into Eq.~\eqref{eq:lambda-R} gives
\begin{equation}
\lambda^2
=
\frac{\mug}{2\rg^2}
\left(2\mug-\rg^2\mu''_\gamma\right).
\label{eq:lambda-direct}
\end{equation}
Since $b_\gamma^2=\rg^2/\mug$, the normalized instability is obtained directly as
\begin{equation}
\eta_\gamma^2
=
\mug-\frac{\rg^2}{2}\mu''_\gamma.
\label{eq:eta-direct}
\end{equation}

For the five-dimensional charged solution, we differentiate Eq.~\eqref{eq:charged-metric} twice and simplify at the orbit using
\begin{equation}
\rg^2\Xi_\gamma
=
2m-\frac{3q^2}{\rg^2}.
\end{equation}
The result is
\begin{equation}
\eta_{\gamma,\mathrm{direct}}^2
=
\frac{2}{\Xi_\gamma^2}
-
\frac{6q^2}{\rg^6\Xi_\gamma^2}
\left(\rg^2\Xi_\gamma-\frac{2\at}{3}\right).
\label{eq:charged-direct-check}
\end{equation}
This agrees exactly with Eq.~\eqref{eq:charged-lamR}. The agreement checks the matter-dependent coefficients without using the field-equation reduction in Appendix~\ref{app:algebra}.

Third, the vacuum limit $q=0$ gives
\begin{equation}
\rg^2=2\rH\sqrt{\rH^2+\at},
\qquad
\Xi_\gamma=\sqrt{1+\frac{\at}{\rH^2}},
\end{equation}
and therefore
\begin{equation}
\eta_\gamma^2
=
\frac{2}{\Xi_\gamma^2}
=
\frac{2\rH^2}{\rH^2+\at}.
\end{equation}
Together with the value of $\mug$ in Eq.~\eqref{eq:5d-orbit-data}, the relation $\lambda^2=\mug\eta_\gamma^2/\rg^2$ reproduces Eq.~\eqref{eq:5d-lambda}. The normalized instability likewise reproduces Eq.~\eqref{eq:5d-lamR}. These checks confirm the Einstein limit, the charged matter contribution, and vacuum saturation.

\section{Derivations of the optical and tidal identities}
\label{app:interpretation}

The physical interpretation uses geometrical identities that can be checked independently of the EGB field equations. Let $A$ and $\mu$ be arbitrary smooth positive functions near a circular null orbit. The orbit condition is $A'_\gamma=2A_\gamma/\rg$, and the coordinate-time instability follows from the radial potential as
\begin{equation}
 \lambda^2=\frac{\mug}{2}
 \left(\frac{2A_\gamma}{\rg^2}-A''_\gamma\right).
 \label{eq:lambda-general-check}
\end{equation}
To obtain the optical curvature, differentiate $b=rA^{-1/2}$:
\begin{equation}
 \frac{\dd b}{\dd r}=A^{-1/2}\left(1-\frac{rA'}{2A}\right).
\end{equation}
This vanishes at the orbit, and a second derivative there gives
\begin{equation}
 \left.\frac{1}{b}\frac{\dd^2b}{\dd r^2}\right|_\gamma
 =\frac{1}{\rg^2}-\frac{A''_\gamma}{2A_\gamma}.
\end{equation}
Since $\dd r/\dd x=\sqrt{A\mu}$, the term containing $\dd b/\dd r$ drops out of the second $x$ derivative. Therefore
\begin{equation}
 \mathcal K_\gamma=-\frac{1}{b_\gamma}
 \left.\frac{\dd^2b}{\dd x^2}\right|_\gamma
 =-\lambda^2,
\end{equation}
confirming Eq.~\eqref{eq:optical-curvature}.

For the spacetime tidal check, the required orthonormal curvature components are
\begin{align}
 R_{\hat r\hat0\hat r\hat0}
 &=\frac{\mu A''}{2A}+\frac{\mu'A'}{4A}
 -\frac{\mu A'^2}{4A^2},\\
 R_{\hat r\hat\phi\hat r\hat\phi}&=-\frac{\mu'}{2r},\\
 R_{\hat a\hat0\hat a\hat0}&=\frac{\mu A'}{2Ar},\\
 R_{\hat a\hat\phi\hat a\hat\phi}&=\frac{1-\mu}{r^2}.
 \label{eq:tidal-components}
\end{align}
Here $\hat a$ is any angular direction perpendicular to the orbit. Mixed time-angular contributions vanish by static spherical symmetry. For $k=e_{\hat0}+e_{\hat\phi}$, summing the first two components and imposing the orbit condition gives
\begin{equation}
 \mathcal E_{\hat r\hat r}
 =\frac{\mug A''_\gamma}{2A_\gamma}-\frac{\mug}{\rg^2}
 =-\frac{\lambda^2}{A_\gamma}.
\end{equation}
Summing the last two gives $\mathcal E_{\hat a\hat a}=1/\rg^2$. Moreover, $\nabla_k e_{\hat r}=(\sqrt{\mug}/\rg)k$ at the orbit, so its screen projection is parallel transported. An affine tangent with the chosen normalization has $\dd t/\dd s=A_\gamma^{-1/2}$; the negative radial eigenvalue thus reproduces the coordinate-time rate $\lambda$. Their trace proves Eq.~\eqref{eq:Ricci-orbit}.

A separate Ricci-component check uses
\begin{align}
 \frac{R_{tt}}{A}&=\frac{\mu A''}{2A}
 -\frac{\mu A'^2}{4A^2}+\frac{\mu'A'}{4A}
 +\frac{(n-2)\mu A'}{2Ar},\\
 R_{\hat\phi\hat\phi}&=\frac{(n-3)(1-\mu)}{r^2}
 -\frac{\mu'}{2r}-\frac{\mu A'}{2Ar}.
\end{align}
Adding these and setting $A'=2A/r$ cancels $\mu'$ and yields Eq.~\eqref{eq:Ricci-orbit} directly. Substituting the EGB identity~\eqref{eq:shadow-exact} then yields Eq.~\eqref{eq:EGB-convergence}; no modified Raychaudhuri equation has been assumed.

For the density check, $\mathfrak m=r^{n-1}W$ and $W=\psi+\at\psi^2$ imply $\Xi^2=1+4\at\mathfrak m/r^{n-1}$. Hence
\begin{align}
 B-(n-3)
 &=\frac{32\pi\at\rho}{(n-2)\Xi^2}
 -\frac{2(n-1)\at\mathfrak m}{r^{n-1}\Xi^2}\nonumber\\
 &=\frac{2\at[r\mathfrak m'-(n-1)\mathfrak m]}
 {r^{n-1}\Xi^2}
 =\frac{r\Xi'}{\Xi}.
\end{align}
This verifies both the sign and the factor $32\pi/(n-2)$ in the density-contrast bound. In units $G=c=1$, $\rho$ and $\bar\rho$ have dimensions of inverse length squared, so $\at\rho$, $B$ and $\eta_\gamma$ are dimensionless, whereas the null-convergence and optical-curvature expressions have dimensions of inverse length squared.

\clearpage
\bibliographystyle{apsrev4-2}
\bibliography{EGB_Lyapunov_Bounds_Revised}
\end{document}